\documentclass[aps,prx,twocolumn,superscriptaddress]{revtex4-2}

\usepackage[compat=1.0.0]{tikz-feynman}
\usepackage{amsmath,amssymb}
\usepackage{multirow}
\usepackage{bm}
\usepackage{changes}
\usepackage{graphicx}
\usepackage{color,bbold,keyval,url,latexsym}
\usepackage{xcolor}
\usepackage{epstopdf}
\usepackage{subfigure}
\usepackage{epsfig}
\usepackage{amsfonts}
\usepackage{mathrsfs}
\usepackage{braket}
\usepackage[toc,page,title,titletoc,header]{appendix}
\usepackage{dsfont,amsthm,amsbsy}
\usepackage[colorlinks,linkcolor=blue,citecolor=blue,anchorcolor=blue]{hyperref}

\usetikzlibrary{quotes,angles}
\newcommand{\obtuseangle}{\kern.08em
\begin{tikzpicture}
    \draw coordinate (a) at (0.14,0);
    \draw coordinate (b) at (0,0);
    \draw coordinate (c) at (-.12,0.18);
    \draw (a) -- (b) -- (c) pic [draw=black]{} ;
\end{tikzpicture}%
\kern.08em%
}

\newcommand{\be}{\begin{eqnarray}}
\newcommand{\ee}{\end{eqnarray}}

\begin{document}

\title{Competition between charge-density-wave and superconducting orders on eight-leg square Hubbard cylinders}
\author{Hong-Chen Jiang}
\email{hcjiang@stanford.edu}
 \affiliation{Stanford Institute for Materials and Energy Sciences,
SLAC National Accelerator Laboratory, 2575 Sand Hill Road, Menlo Park, CA 94025, USA}
\author{Thomas P. Devereaux}
\email{tpd@stanford.edu}
\affiliation{Stanford Institute for Materials and Energy Sciences,
SLAC National Accelerator Laboratory, 2575 Sand Hill Road, Menlo Park, CA 94025, USA}
\affiliation{
Department of Materials Science and Engineering, Stanford University, Stanford, CA 94305, USA}
\affiliation{
Geballe Laboratory for Advanced Materials, Stanford University, Stanford, CA 94305, USA}
\author{Steven A. Kivelson}
\email{kivelson@stanford.edu}
\affiliation{
Department of Physics, Stanford University, Stanford, CA 94305, USA}

\date{\today}
\begin{abstract}
The issue of whether  $d$-wave superconductivity (SC) 
occurs in the square-lattice Hubbard model with $U$ of order of the bandwidth has been one of the most debated issues to emerge from the study of high temperature SC. Here, we report variational results on  eight-leg cylinders with   next-nearest-neighbor hopping in the range $-0.5 t \leq t'\leq 0.25 t$ with $U = 8t$ and $12t$ and doped hole concentrations $\delta=1/12$ and  $1/8$. For $t'\leq 0$, the ground-state appears to be a charge-density wave (CDW) of one sort or another with SC correlations that are 
extremely short-ranged. In contrast, in some cases, the local magnetic order has a correlation length greater than half the cylinder width - suggestive that magnetic order might also arise in the 2D limit. For  $t'>0$, our results depend more strongly on boundary conditions (periodic vs antiperiodic), making it still harder to correctly guess whether SC or CDW correlations  dominate in the 2D limit. These results were obtained  employing  matrix-product states with bond dimensions large enough that energy
differences as small as  $10^{-3}t$ per site can be resolved.
\end{abstract}
\maketitle

\section{Introduction}%
The Hubbard model on a square lattice with intermediate values of $U=8t-12t$ (i.e. comparable to the band-width, $8t$) is widely believed to capture much of the essential physics of the high temperature superconducting cuprates, despite the fact that it is not a reasonable representation of the local electronic structure of the actual materials. Indeed, the Hubbard model on various lattices is considered to be the paradigmatic model of strongly correlated electron systems more generally, even in the context of materials that are microscopically even further distant from it. In common with the cuprates, the Hubbard model robustly exhibits\cite{annurev,gullrev} an antiferromagnetic insulating ``parent'' state when undoped (one electron per site, $n=1$) and strong local tendencies to $d$-wave pairing and to the various types of stripe order (unidirectional spin and charge density wave order, SDW and CDW) that also appear to be ubiquitous features of at least the hole-doped cuprates \cite{Fradkin2015}. However, while it is clear that for sufficiently small $U/t$, the low temperature ordered phase is adiabatically related to the $d$-wave superconducting state seen in the cuprates \cite{chubukov,Raghu2010}, the issue of whether the ground-state at intermediate $U/t$ is superconducting (SC) at all has been bitterly contested.

Serious numerical approaches to this model - and the closely related $t-J$ model - have reached contradictory conclusions concerning the ground-state phase diagram, depending to some degree on details of the assumed band-structure, i.e. the sign and magnitude of the second-neighbor hopping term $t'$, but also on the way the calculations were carried out.
Specifically, a number \cite{Jiang2018tJ,Jiang2019Hub,Jiang2020YF,Gong2021,Lu2024,Xu2024,Chen2025,Hager2005,Ponsioen2019,Qin2020,Jiang2021,Jiang2022,Jiang2024,Xu2024HubSix,Liu2025,Xu2025Dia,Wang2025EP} of studies have claimed to find robust, $d$-wave SC order while others have concluded that some form of competing order preempts SC. In particular, results obtained via density matrix renormalization group (DMRG) simulations of  4-leg ladders robustly reveal some form of algebraic  
SC order. However, even sticking with DMRG studies, conflicting results have been obtained on  6-leg and wider ladders and cylinders. This is due in part to the difficulty of making sure that the DMRG results have converged to the true ground state on a given cluster. 

In essence, this history of conflicting results reflects the existence of multiple possible ground-states with similar local correlations but with widely divergent long-range order - a situation that has been referred to as ``intertwined orders'' \cite{Fradkin2015,Fernandes2019,Mai2022}.  Indeed, 
various theoretical studies are suggestive that, in the 2D limit, there exist multiple possible ordered states with energy density differences 
$\sim 10^{-2} t$ per site \cite{Corboz2014,Zheng2017,Ponsioen2019,Liu2025}. One approach to mitigating this problem is to increase the bond dimension in order to increase the accuracy of the ground states obtained. Moreover, formulating the problem in a way that incorporates the microscopic symmetries leads to variational ground states that properly respect the $U(1)$ and $SU(2)$ symmetries of the Hubbard Hamiltonian.

\begin{figure}[!th]
\centering
  \includegraphics[width=1\linewidth]{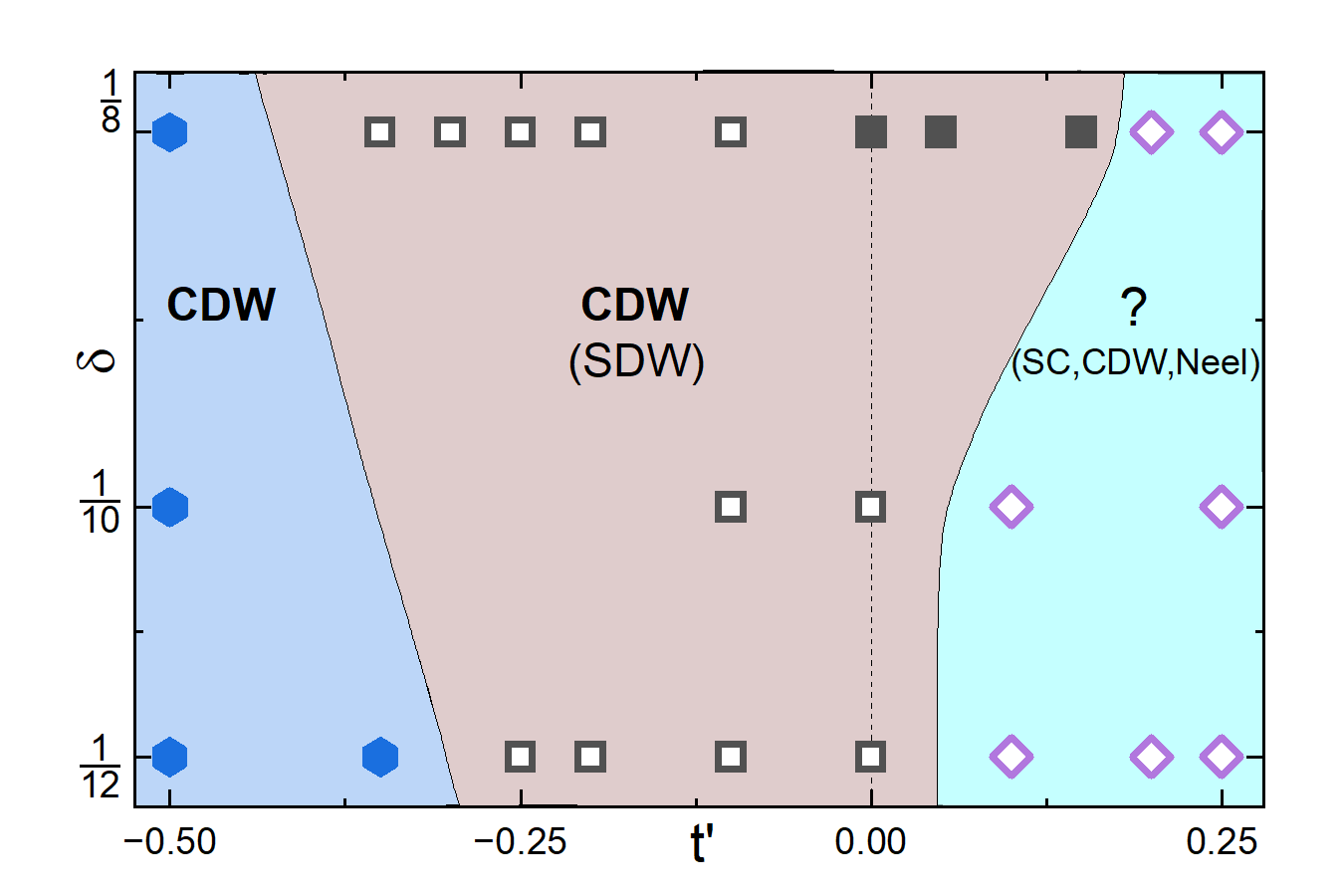}
\caption{Ground state phase diagram of the Hubbard model on an eight-leg cylinder as a function of $t'$ and $\delta$ for $U=12t$. Everywhere except the region marked with a {\bf ?}, the heirarchy of ordering tendencies (if not all details of the order) is independent of BCs. Some form of apparently long-range CDW order arises with substantial amplitude in all regions labeled {\bf CDW}; where CDW appears within parenthesis, CDW order is either much weaker, or fluctuating.  Regions with notable subdominant magnetic correlations (defined as a  correlation length in excess of 4 lattice constants)  are labeled (SDW) when it is locally incommensurate and (N\'eel) where commensurate.  SC correlations are  always locally $d$-wave-like, but decay exponentially with distance; only in the region labeled with SC in parenthesis is  the SC correlation length in excess of 2 lattice constants. Shaded regions are guides to the eye. Open and filled symbols, respectively, indicate points at which the cylinder with ABCs or PBCs has the lower energy, respectively.}
\label{Fig:Phase}
\end{figure}

{\it Broader Context:}  Identifying the ground states of the Hubbard model is of foundational importance.  If in a reasonable range of parameters, the Hubbard model were to prove to be a robust $d$-wave superconductor, it would
suffice for theoretical studies to focus on the mechanism of high-temperature
SC in this relatively simple, no-frills model problem.  If not, this raises the important open question: What is needed in addition to capture the essence of the problem?

 {\it Technical aspects:}
 Here, we report on the results of unprecedentedly extensive DMRG studies, with bond-dimensions of $SU(2)$ multiplets up to $D = 60,000$ (equivalent to $m \approx 196,000$ $U(1)$ states), for the Hubbard model on cylinders of width $W=8$ and lengths up to $L=36$, i.e. with $N= W \times L = 288$ sites. Specifically, we impose open boundary conditions at the cylinder ends, and either periodic (PBC) or anti-periodic (ABC) boundary conditions around the cylinder corresponding to 0 or $1/2$ a magnetic flux quantum through the cylinder. (We also studied 8-leg ladders with open boundary conditions in both directions, as reported in the Supplemental Materials (SM).) The large bond-dimension proved {\it absolutely} necessary to obtain clearly converged results (e.g., with respect to spin-rotational and other symmetries), so the fact that the nature of the observed ground-state at large $D$ or $m$ is in some cases qualitatively different than that observed for the smaller values typical of previous studies, suggests that earlier results (especially on relatively wide ladders and cylinders) need to be viewed with care.

{\it Global phase diagram:} Some features of the ground-state associated with these systems (especially, the precise form of the observed CDW order) are found to depend on boundary conditions, which a priori implies a degree of uncertainty about what they teach us about the system in the thermodynamic limit ($N$ and $W\to \infty$).  However, essential aspects of the results are boundary condition independent, especially for $t'\leq 0$ - in particular that there appears to be long-range CDW order and that the SC correlations fall exponentially with a short correlation length. This encourages us to propose that the global ground-state phase diagram of the 8-leg cylinder, shown in Fig. 1 for $U=12t$, is indicative of the  behavior to be expected in the thermodynamic limit.  Here $\delta$ is the ``doping concentration'' ($n=1-\delta$) and $t'$ is the second-neighbor hopping matrix element. 

\begin{figure*}
  \includegraphics[width=1.0\linewidth]{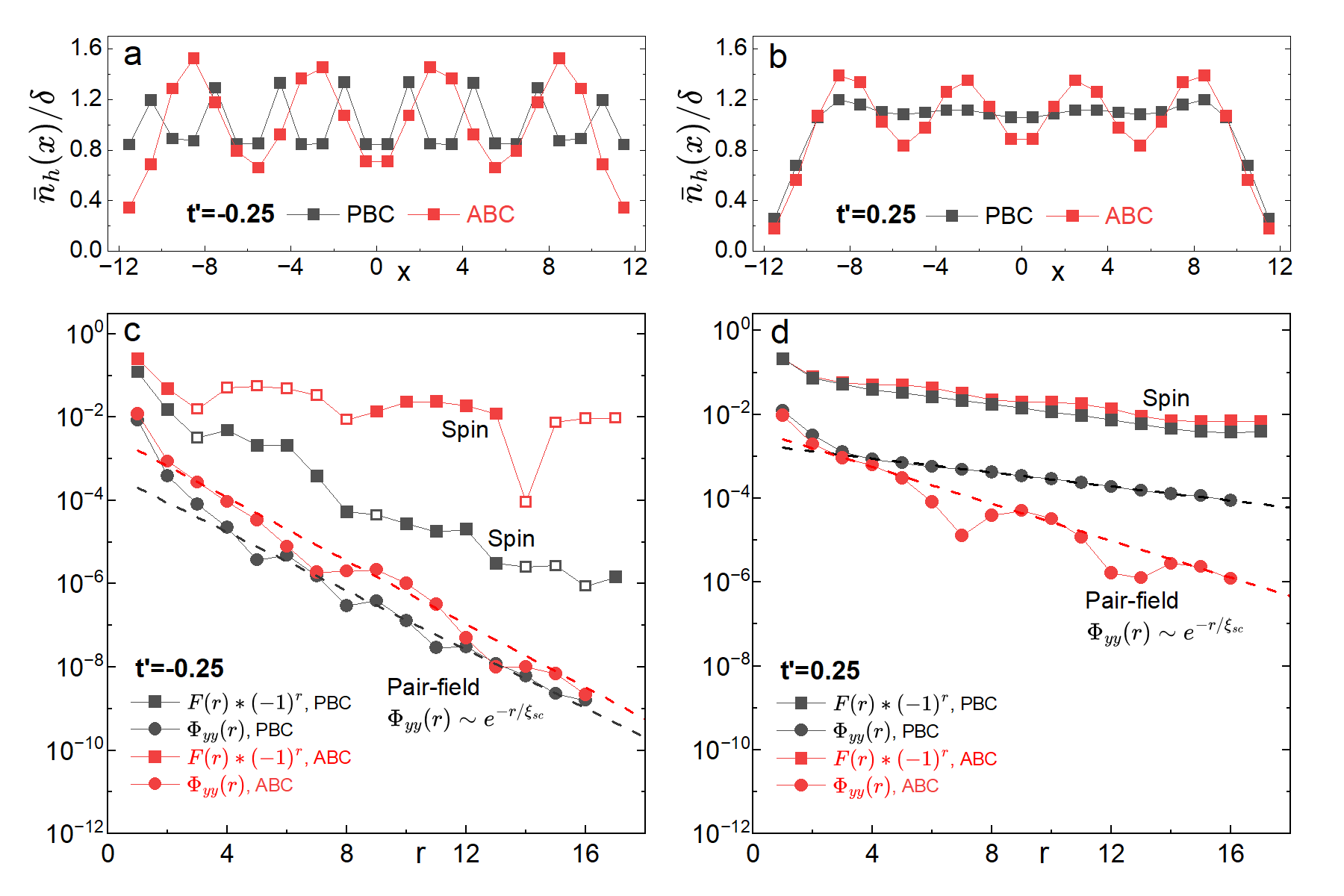}
\caption{Ground state correlations for $N=24\times 8$ Hubbard cylinders with $t'=-0.25$ (panels a \& c) and $t'=+0.25$ (panels b \& d), $U=12t$ and $\delta=1/12$. Panels a and b show the rescaled doped hole  density, $\bar n_h(x)/\delta$. Panels c and d show the magnitude of the two-point correlations on a log-linear plot as a function of distance $r$ along the cylinders: $F(r)*(-1)^r$ is the staggered spin-spin correlation function  and  $\Phi_{yy}(r)$ is the pair-field correlation function for singlet, nearest-neighbor pairs oriented perpendicular to the cylinder, where dashed lines denote exponential fits $\Phi_{yy}(r)\sim e^{-r/\xi_{sc}}$ with correlation length $\xi_{sc}$. The sign structure of the spin correlations is indicated by the full symbols  (where positive) and open symbols (where negative).  The  N\'eel-like character of the spin correlations for $t'>0$ is reflected in the absence of open symbols, while a pattern of anti-phase domain walls for $t'<0$ can be seen from the alternating regions of open and closed symbols.}
\label{Fig:NxCor}
\end{figure*}

The orders indicated in parentheses in the figure refer to significant short-range orders, i.e. those with correlations that decay   with distance (apparently exponentially as shown, for example, in Fig. \ref{Fig:NxCor}) but with correlation lengths that are not too short. Specifically, (N\'eel) refers to spin correlations that decay with a correlation length that is approximately 4 lattice constants or longer, and which show sign oscillations corresponding to an ordering vector $\vec Q = (\pi,\pi)$, while (SDW) refers to similarly decaying stripe-like correlations with $\vec Q =(\pi+\delta q,\pi)$. In most of the phase diagram, the SC correlations fall with a correlation length measurably shorter than 2 lattice constants, except in the portion of the positive $t'$ region marked (SC), where the SC correlation length can be as large as 4 lattice constants for PBCs and is at least 2 for ABCs. (Correlation lengths for various values of $t'/t$ are shown in Table \ref{Tab:One} and Table \ref{Tab:Sone} in the SM.)

The most significant feature of Fig. \ref{Fig:Phase} is that for $t'\leq 0$ (hole-doping), independent of BCs, the dominant long-range correlations involve the CDW order. Spin order is sub-leading. The SC correlations have a $d$-wave like sign structure, but are extremely short-range correlated. For $t'>0$, (electron doping), the conclusions are less clear. Here we encounter a strong dependence on boundary conditions, finding substantial variations in correlations lengths and CDW amplitudes. In particular, for PBCs and $t'>0$, we find long enough SC correlation lengths that SC in the 2D limit seems plausible, in agreement with recent iPEPs \cite{Zhang2025} and determinant quantum Monte Carlo (DQMC) \cite{Wang2025} studies.   Our methods and how we arrived at our conclusions are summarized in the following sections.

\section{Model and Method}\label{Sec:Method}
We employ DMRG \cite{White1992,White1993} to study the ground state properties of the Hubbard model on the square lattice, which is defined by the Hamiltonian
\begin{eqnarray}\label{Eq:Ham}
H=-\sum_{ij,\sigma} t_{ij} \left(\hat{c}^\dagger_{i\sigma} \hat{c}_{j\sigma} + h.c.\right) + U\sum_i\hat{n}_{i\uparrow}\hat{n}_{i\downarrow}.
\end{eqnarray}
Here $\hat{c}^\dagger_{i\sigma}$ ($\hat{c}_{i\sigma}$) is the electron creation (annihilation) operator on site $i=(x_i,y_i)$ with spin polarization $\sigma$, $\hat{n}_{i\sigma}$ is the electron number operator, $t$ and $t'$  are, respectively, the nearest-neighbor (NN) and next-nearest-neighbor (NNN) hoppings, and $U>0$ is the on-site Coulomb repulsion. The total number of electrons is $N_{el}$, hence the doping level (relative to the half-filled band) is $\delta\equiv (N-N_{el})/N$. For the present study, we consider parameters in the range $-0.50t\leq t'\leq 0.25t$, $U=12t$ and doping concentrations $\delta=1/12\ \& \ 1/8$. In some cases, we have also studied the problem with $U=8t$, for which qualitatively similar results have been found, as discussed in the SM. (A particle-hole transformation changes the signs of $t'$ and $\delta$, which motivates referring to the results with $\delta >0$ and $t'>0$ and $t'<0$ as being comparable to ``electron-doped'' and ``hole-doped'' cuprates, respectively.) 

We have implemented a version of DMRG \cite{McCulloch2002} that respects the $SU(2) \equiv SU(2)_{spin}\otimes U(1)_{charge}$ symmetry of the model, i.e., full spin rotational symmetry and charge conservation. We perform around 240 sweeps and keep bond dimensions of $SU(2)$ multiplets up to $D$=60,000. This ensures accurate results with a typical truncation error $\epsilon\approx 5\times 10^{-6}$. More details are provided in the SM.

\section{Results}
To characterize the properties of the cylinders, we have computed various ground-state correlation functions. The position dependent doped hole density $n_h(\vec R)$ and the equal time spin and pair-field two-point correlation functions, defined here, are shown in Fig. \ref{Fig:NxCor} for characteristic values of the parameters:
\begin{eqnarray}
n_h({\bf R}) = 1-\langle  n({\bf R})\rangle  \label{Eq:nh}
\end{eqnarray}
\begin{eqnarray}
\Phi_{\alpha\beta}({\bf R};{\bf R}')=\langle\Delta^{\dagger}_{\alpha}({\bf R})\Delta_{\beta}({\bf R}')\rangle. \label{Eq:SC}
\end{eqnarray}
\begin{eqnarray}\label{Eq:SS}
    F({\bf R};{\bf R}')=\langle \vec{S}({\bf R})\cdot \vec{S}({\bf R}')\rangle.
\end{eqnarray}
where $n({\bf R})\equiv \sum_\sigma c^\dagger_{{\bf R},\sigma} c_{{\bf R},\sigma}$ is the electron density operator at the site ${\bf R}=(x,y)$,  the spin-singlet pair creation operator on a bond in the direction $\hat e_\alpha=\hat{x}$ or $\hat{y}$ is 
$\Delta^{\dagger}_{\alpha}({\bf R})\equiv \frac{1}{\sqrt{2}}[\hat{c}^{\dagger}_{{\bf R},\uparrow}\hat{c}^{\dagger}_{{\bf R}+\hat e_\alpha,\downarrow}+\hat{c}^{\dagger}_{{\bf R}+\hat e_\alpha,\uparrow}\hat{c}^{\dagger}_{{\bf R},\downarrow}]$, 
and $\vec S({\bf R}) \equiv \sum_{\sigma,\sigma'} \ c^\dagger_{{\bf R},\sigma}\vec \tau_{\sigma,\sigma'}c^\dagger_{{\bf R},\sigma'}$ is the spin operator.

In the absence of translation symmetry breaking, the two-point correlation functions must depend only on the distance between sites. To explore this intrinsic behavior, we define the correlators as a function of separation along the long (x) axis of the cylinder, 

\begin{eqnarray}
&&\Phi_{\alpha\beta}(r)\equiv \Phi_{\alpha\beta}({\bf R}_0;{\bf R}_0+r\ \hat{{\bf x}})\nonumber \\
&& F(r)\equiv F({\bf R}_0;{\bf R}_0+r\ \hat{{\bf x}}).
\end{eqnarray}
where ${\bf R}_0= (x_0,y)$ is a reference point taken as $x_0\approx L/4$ to minimize the open boundary effect and $r$ is the displacement between sites or bonds in the $\hat{x}$ direction.

In Fig. \ref{Fig:NxCor} we show the behavior of these correlation functions for $U=12t$, $t'=\pm 0.25 t$ and $\delta= 1/12$ under different boundary conditions. Our decision to highlight $\delta=1/12$ rather than $\delta = 1/8$ was motivated by several considerations:
We are particularly interested in the consequences of lightly doping a Mott insulator. This choice minimizes commensurability effects (discussed below) associated with the period of the dominant ``half-filled stripes'' and $W=8$ around the cylinder. It also lets us exhibit the case ($\delta = 1/12$, $t' = +0.25t$ and PBCs) that has the longest-range SC correlations among any of the parameter sets we have explored.  However, curiously, this is also the case that required the largest bond dimension to achieve convergence of the charge density, $n(\vec R)$.

\begin{figure*}
  \includegraphics[width=1.0\linewidth]{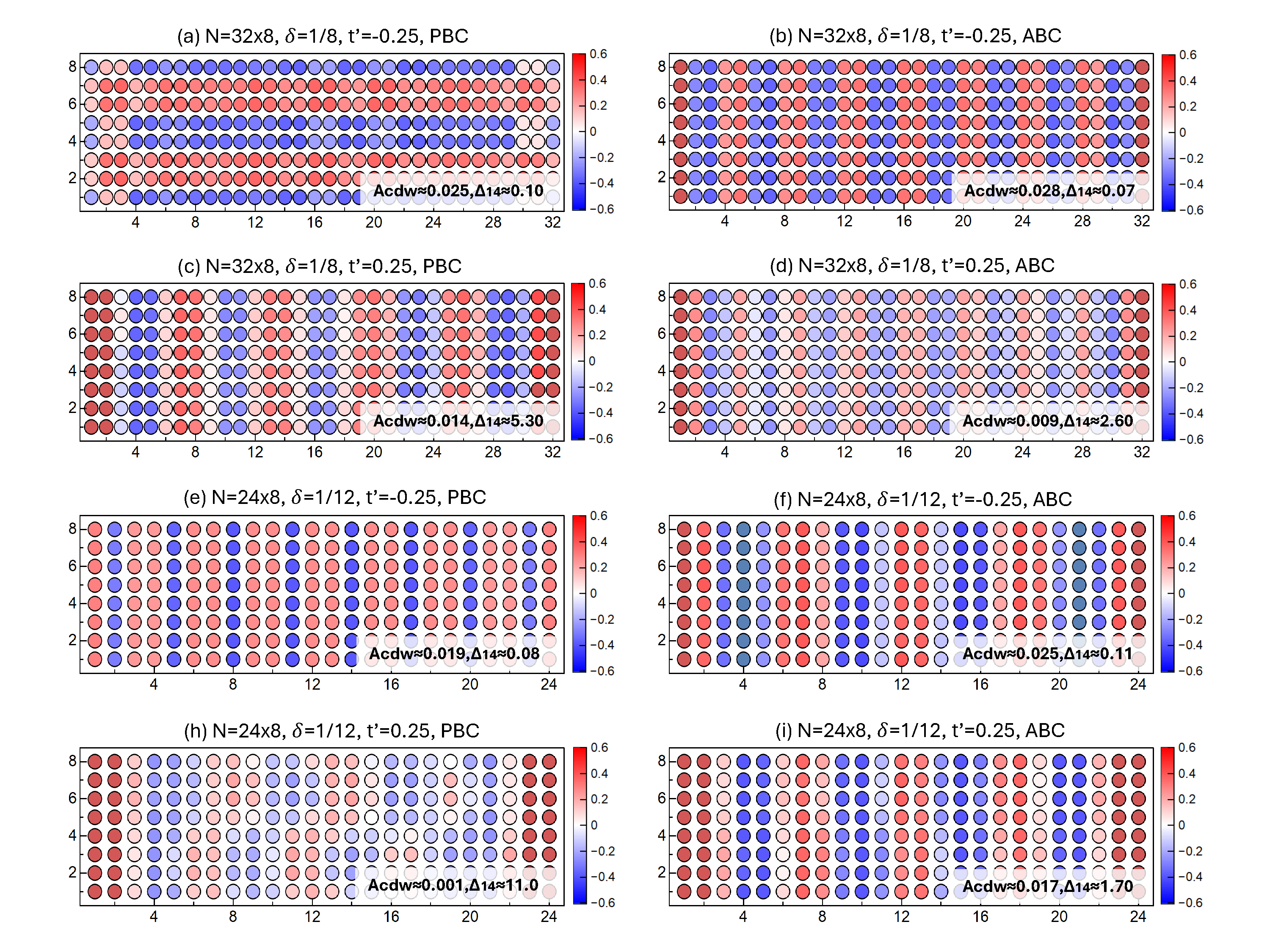}
\caption{(Color online) Rescaled hole density profiles $dn_h(x,y)\equiv{\rm sign}[\delta n_h(x,y)]\times \sqrt{|\delta n_h(x,y)|}$ for 8-leg Hubbard cylinders with $U=12t$ and $t'=\pm 0.25t$, shown at $\delta=1/8$ (panels a - d) and $\delta=1/12$ (panels e - i). Here, $\delta n_h(x,y)=(n_h(x,y)-\overline{n_h})/\delta$, $\overline{n_h}$ is the average hole density and 
$A_{cdw}$ is the CDW amplitude (defined in Eq. \ref{Eq:Acdw}) and $\Delta_{14}$ is a measure of the SC correlations (defined in Eq. \ref{Delta}). Results with PBCs are exhibited in the right column, and ABCs in the right. The sites in dark red and blue indicate values outside the plotted scale.}
\label{Fig:Nxy}
\end{figure*}

The upper panels of Fig. \ref{Fig:NxCor} show the hole density, 
\begin{equation}
  \bar n_h(x)\equiv W^{-1} \sum_y n_h(x,y), 
\end{equation}
averaged over rungs of the cylinder, as a function of position along the cylinder. Since for this set of parameters, the CDW pattern (shown in Fig. \ref{Fig:Nxy}e-i) is essentially unidirectional (stripes around the cylinder), this averaging procedure is a useful method of data compression. Although generically one expects some form of density modulations as one approaches the edge of a 1D system, the fact that where the CDW amplitude is large (especially for $t'\leq 0$), the amplitude of the density modulations are essentially $x$ independent,  is strongly indicative of the existence of long-range CDW order in the $L \to \infty$ limit.

For $t'<0$, panel 2a, CDW order with a substantial magnitude is present for either boundary condition.  
The order shown is commensurate for ABC, with period 6 for $\delta=1/12$ - in the language of stripe order, these are referred to as ``half-filled stripes.''  However, other sorts of commensurate CDW order can be observed for different boundary conditions (such as period 3 for PBC here).

For $t'>0$ on the other hand, different boundary conditions yield strongly different results. For PBC, Fig. \ref{Fig:NxCor}b, only small modulations of hole density is observed away from the ends of the cylinder, while for ABC, strong density modulations with the same amplitude and period as for $t'<0$ are observed. This strong boundary condition dependence underlies our uncertainty about the phase diagram in this region, as indicated in Fig. \ref{Fig:Phase}.

A strong dependence on boundary conditions is also manifest in the spin and superconducting correlations. The lower panels of Fig. \ref{Fig:NxCor} show the spin correlations and the pair-field correlations corresponding to the pair-creation on $y$-directed bonds, $\Phi_{yy}$. More results for $\Phi_{yy}$ and $\Phi_{xx}$ are presented in the SM. These correlation functions are plotted on a log-linear scale, making manifest that over the accessible range of distances, they decay exponentially with distance, with correlation lengths that are sensitive both to the sign of $t'$ as well as boundary conditions. 

It is clear from Fig. \ref{Fig:NxCor}c that for $t'<0$, the SC correlations are weak and  short-ranged, independent of BCs, with $\xi_{sc} \lesssim 2$. Moreover, the spin correlations are more or less consistent with a pattern of stripes with twice the period of the CDW, but with correlation lengths (listed in Table \ref{Tab:One}) that differ by a factor 5 or so  dependent on BCs: the stripes are more prominent for ABC than PBC. It is notable that the strength of the CDW and SC orders appear, in this case, to be largely uncorrelated with the strength of the spin correlations. Despite the BC dependence of the spin correlations, the approximate  BC  independence of  the amplitude of the CDW order and the rapid decay of the SC correlations strongly support our identification of this as a non-superconducting CDW phase in Fig. \ref{Fig:Phase}.

For $t'>0$, Fig. \ref{Fig:NxCor}d, we see that the spin correlations are N\'eel like and not very sensitive to BCs, despite the strong BC-dependent variation in the CDW amplitudes. The SC correlations are substantially larger, by 2-3 orders of magnitude  at accessible distances, than for $t'<0$, and also substantially BC-dependent. Specifically, stronger SC correlations are seen for PBCs, where the SC  correlation length is comparable to the spin, $\xi_{SC}\sim\xi_s\gtrsim 4$, while the CDW amplitude is  relatively weak. Conversely, for ABCs, the SC correlations are noticeably weaker, although still stronger than for $t'<0$. The fact that even the relative strength of the different orders is BC-dependent is the origin of our uncertainty in identifying phases in the corresponding portions of Fig. \ref{Fig:Phase}. However, what is clear is that, given that the anticorrelation between the CDW and SC correlations, there is substantial competition between CDW and SC orders.

\begin{table}[]
  \centering
  \begin{tabular}{|c|c|c||c|c|c|c|}\hline
  $N$ &$\delta$  $\ \ $ &$t'/t$  $\ \ $ & $ A_{cdw}$  $\ \ $ & $\xi_{spin}$  $\ \ $ & $\xi_{SC}$  $\ \ $ &  $\delta\varepsilon \times 10^{3}$\\
\hline\hline\hline
  $32\times 8$ &$ 1/8$ & $-0.25$ & 0.028 & 5.6 & 1.1 & \\ 
   &  &  & (0.025) & (3.2) & (1.5) & +2.56$t$ \\  
\hline
  & & $ $ $0$ & 0.024 & 5.9 & 1.3 & \\
  &  &  & (0.033) & (4.2) & (1.4) & -1.43$t$ \\  
\hline
  & $ $ & $ 0.25$ & 0.009 & 5.5 & 2.2 & \\ 
  &  &  & (0.014) & (3.0) & (2.9) & +4.37$t$ \\ 
\hline\hline
    $24\times 8$ & $ 1/12$ & $ -0.25$ &  0.025 & 6.9 & 1.2 & \\ 
  &  &  &  (0.019) & (1.5) & (1.3) & +1.00$t$ \\ 
\hline
  & & $ $ 0 & 0.030 & 7.4 & 1.2 & \\
  &  &  & (0.025) & (5.9) & (1.4) & +0.79$t$ \\ 
\hline
  & $ $ & $0.25$ & 0.017 & 6.2 &  2.1 & \\
  &  &  & (0.001) & (4.8) & (5.2) & +2.81$t$ \\ 
\hline\hline
  $36\times 8$ & $1/12$ & $ -0.25$ & 0.026 & 6.6 & 1.1 & \\ 
  &  &  & (0.020) & (1.3) & (1.0) & +0.83$t$ \\  
\hline
  & $ $ & $0$ & 0.030 & 8.2 & 1.2  & \\ 
  &  &  & (0.025) & (5.7) & (1.3) & +0.75$t$ \\ 
\hline\hline
 \end{tabular}
\caption{Representative results for $U=12t$ and system sizes $N=L\times W$, where for each set of parameters, the upper entry is for ABCs and the lower entry (in parenthesis) is for PBCs. $\xi_{spin}$ and $\xi_{SC}$ are, respectively, the spin and SC correlation lengths. $\delta\varepsilon$ is the difference in the ground-state energy density per site of the system with PBC- ABC - i.e. ABC has lower energy when $\delta\varepsilon >0$. (Results for other parameters are summarized in Table \ref{Tab:Sone} in SM.)}
\label{Tab:One}
\end{table}

The real-space charge-density profile is shown in Fig. \ref{Fig:Nxy} for various sized clusters, for $\delta=1/8 \ \& \ 1/12$, $t'/t=\pm 0.25$, and for both  PBCs and ABCs. (We use a square-root scaling of the color scale to be able to visualize the CDW order across a range of strengths.) The observed CDW varies in its orientation, magnitude, and period, suggesting a near degeneracy of various types of CDW ordering patterns. For instance, comparing Figs. \ref{Fig:Nxy} a and b, one sees that for $\delta=1/8$ and $t'=-0.25t$, the same period stripes arise independent of boundary conditions, but that for PBCs they are oriented along the cylinder and for ABCs they are perpendicular to this axis. (This strong BC-dependence is likely related to the commensurability between the CDW order and the cylinder width, $W$. It is interesting to note that similar horizontal half-filled stripes were also observed\cite{Gu2025ML}  using a new - and very promising - variational approach to the same model with nearly the same parameters.)  It is perhaps unsurprising that - given the relatively long period of the CDW order (typically period 4 for $\delta=1/8$ and period 6 for $\delta=1/12$) - that there is still a BC dependence even for relatively wide cylinders.

In the lower right corner of each panel of Fig. \ref{Fig:Nxy} (and also in the Tables) we characterize  the strength of the CDW order by the root-mean-squared variance in $n_h(\vec R)$,
\begin{equation}\label{Eq:Acdw}
   A_{cdw}\equiv \sqrt{{N'}^{-1}\sum_{xy} [n_h(x,y) - \overline{n_h}]^2}
\end{equation}
averaged over the $N' = N/2$ sites nearest the middle of the cylinder. (Because there is some tendency for holes to concentrate near the end of the cylinder, the average doped hole density in this region, $\overline{n_h}$, is typically slightly less than $\delta$.) To get a feeling for magnitudes, a ``maximal'' CDW in which $n_h=0$ on half the sites and $n_h=2\delta$ on the other half would have $A_{cdw}=|\delta|$;  we interpret the cases in which the observed CDW order is 10\% or more of this as having ``strong'' CDW order. Also in Fig. \ref{Fig:Nxy}, we represent the strength of the SC correlations by the value of the pair-field correlation function at an arbitrarily chosen (longish) distance, 
\begin{equation}
    \Delta_{14}\equiv \sqrt{|\Phi_{yy}(14)|}\times 10^3.
    \label{Delta}
\end{equation}

As can bee seen in panels a, b, e, and f of Fig. \ref{Fig:Nxy}, and for a broader range of parameters in Table \ref{Tab:One} (and \ref{Tab:Sone} in the SM), the CDW correlations are strong and the SC correlation length, $\xi_{sc} \leq 2$, independent of BCs, for all $t'\leq 0$. There is a general tendency (although not a perfect correlation) that the stronger the CDW the weaker the SC. This is broadly reminiscent of the duality between these two orders in the Luther-Emery phases seen in 4-leg ladders \cite{Jiang2018tJ,Jiang2019Hub,Jiang2020YF}, although in the present case neither order appears to be power law correlated. Moreover, given the variation across boundary conditions, doping and cluster sizes, the long-distance asymptotics  may be more complex and deserves further scrutiny. 

An important quantitative measure of the sensitivity to BCs is the difference between the ground state energy densities for PBC and ABC, $\delta\varepsilon\equiv [E_{PBC} - E_{ABC}]/LW$. That these differences are small, $\delta\varepsilon\sim 10^{-3}t$, supports the supposition that $W=8$ is large enough to capture much of the physics of the 2D system. A scan of the results in Table I (and Table \ref{Tab:Sone} in SM) shows that for most parameter sets, the energy is lower for ABCs, which also tends to show stronger CDW order and weaker SC correlations.

\section{Summary and future considerations}%
DMRG provides an extremely flexible variational approximation to the ground-state, and so when applied to finite size systems, is guaranteed to converge to the exact result for large enough bond dimensions. On this basis, we believe we have convincingly shown that for the range of parameters studied, the ground-state of the 8-leg Hubbard model with $t'\leq 0$ and $0 < \delta \leq 1/8$ is dominated by CDW order, with exponentially decaying SDW correlations, and SC correlations that typically decay with a significantly smaller correlation length than the SDW. These qualitative conclusions pertain for both PBCs and ABCs. For $t'>0$ on the other hand, although N\'eel spin correlations are always present, depending on BCs, either strong or weak CDW order and correspondingly either weak or strong SC correlations are found.

It is, however, legitimate to question whether these results reflect the properties of the model in the 2D limit. We have carried out calculations for different values of $L$: specifically $L=16$ \& 28 \& 32 for $\delta=1/8$ and $L=24\ \&\ 36$ for $\delta=1/12$. That the results are largely the same for these different, all quite large values of $L$ is compelling evidence that the results persist for $L \to \infty$ cylinders. The justification for extrapolating our results to larger cylinder circumference, $W$, is less conclusive. However, the differences between the results with PBC and ABC provides a plausible metric of the finite size effects. While there are BC-dependent differences in the observed properties, the fact that the hierarchy of orders is largely independent of BCs for $t'\leq 0$ is the strongest evidence we have that in this range of parameters, the results for $W=8$ are representative of the $W\to\infty$ limit. The finite $W$ effects appear to be more qualitatively important for $t'>0$.

Concerning the differences in conclusions reached in prior DMRG studies - we believe that earlier results were generally carried out with insufficiently large bond dimension. Moreover, convergence to the true ground-state  is clearer when the full $SU(2)$ spin-rotational symmetry and the $U(1)$ charge conservation symmetry of the model are explicitly respected.

It has been widely noted that the various orders that are most prominent in short-distance correlations of the Hubbard model are also those that are most widely seen in the cuprates. Similarly, much of the high temperature, normal state properties of the Hubbard model at elevated temperatures extracted from DQMC and METTS calculations exhibit features that are highly reminiscent of the analogous properties in the cuprates \cite{Huang2021,Wietek2021}. All this suggests that the Hubbard model likely incorporates much of the essential physics of the cuprates.  

This, combined with our present results, raises the important question - what relatively minor modifications of the Hubbard model are needed to obtain robust, $d$-wave SC? Among many possibilities, a few that we hope to pursue in the future are:
1) For somewhat larger $\delta$ than we have explored, CDW may be  less strong while correlation effects remain  strong enough that high temperature SC is possible.
2) Since the Hubbard model is superconducting in the small $U$ limit, the present results imply the existence of a critical  $U_c < 8t$ such that it is a superconductor for $U<U_c$.  However, since for small $U$, $T_c$ is parametrically small, $T_c \sim \exp[-\alpha (8t/U)^2]$ where $\alpha$ is a number of order one \cite{Raghu2010,annurev}, unless $U_c$ is close to $8t$, this still leaves open the problem of the origin of ``high temperature SC.'' 
3) SC might well be enhanced by the addition of a nearest-neighbor attractive $V$ that could come from integrating out suitable sorts of optical phonons \cite{Chen2021,Peng2023,Xu2024HubSix}.
4) In the  hole-doped cuprates, there is  more than one important electronic orbital per unit cell \cite{Emery1987}; this additional microscopic complexity could play an essential role in obtaining robust, high temperature  SC.
5) Motivated by the fact that strong enhancement of long-range SC correlations has been found in previous studies of 6-leg ``striped'' Hubbard cylinders \cite{Jiang2022Stripe}, it is worth exploring models with suitably engineered patterns of inhomogeneous couplings.

\acknowledgements
We acknowledge  discussions with Dong-Ning Sheng, Zheng-Yu Weng, Yi-Fan Jiang, Hong Yao, Antoine Georges, and Shiwei Zhang. This work was supported by the Department of Energy, Office of Science, Basic Energy Sciences, Materials Sciences and Engineering Division, under Contract DE-AC02-76SF00515. The data used to generate the figures are deposited in 
Ref.~\footnote{\href{https://doi.org/10.6084/m9.figshare.30689384}{Figshare dataset: 10.6084/m9.figshare.30689384}}.


\begin{thebibliography}{39}%
\makeatletter
\providecommand \@ifxundefined [1]{%
 \@ifx{#1\undefined}
}%
\providecommand \@ifnum [1]{%
 \ifnum #1\expandafter \@firstoftwo
 \else \expandafter \@secondoftwo
 \fi
}%
\providecommand \@ifx [1]{%
 \ifx #1\expandafter \@firstoftwo
 \else \expandafter \@secondoftwo
 \fi
}%
\providecommand \natexlab [1]{#1}%
\providecommand \enquote  [1]{``#1''}%
\providecommand \bibnamefont  [1]{#1}%
\providecommand \bibfnamefont [1]{#1}%
\providecommand \citenamefont [1]{#1}%
\providecommand \href@noop [0]{\@secondoftwo}%
\providecommand \href [0]{\begingroup \@sanitize@url \@href}%
\providecommand \@href[1]{\@@startlink{#1}\@@href}%
\providecommand \@@href[1]{\endgroup#1\@@endlink}%
\providecommand \@sanitize@url [0]{\catcode `\\12\catcode `\$12\catcode `\&12\catcode `\#12\catcode `\^12\catcode `\_12\catcode `\%12\relax}%
\providecommand \@@startlink[1]{}%
\providecommand \@@endlink[0]{}%
\providecommand \url  [0]{\begingroup\@sanitize@url \@url }%
\providecommand \@url [1]{\endgroup\@href {#1}{\urlprefix }}%
\providecommand \urlprefix  [0]{URL }%
\providecommand \Eprint [0]{\href }%
\providecommand \doibase [0]{https://doi.org/}%
\providecommand \selectlanguage [0]{\@gobble}%
\providecommand \bibinfo  [0]{\@secondoftwo}%
\providecommand \bibfield  [0]{\@secondoftwo}%
\providecommand \translation [1]{[#1]}%
\providecommand \BibitemOpen [0]{}%
\providecommand \bibitemStop [0]{}%
\providecommand \bibitemNoStop [0]{.\EOS\space}%
\providecommand \EOS [0]{\spacefactor3000\relax}%
\providecommand \BibitemShut  [1]{\csname bibitem#1\endcsname}%
\let\auto@bib@innerbib\@empty
\bibitem [{\citenamefont {Arovas}\ \emph {et~al.}(2022)\citenamefont {Arovas}, \citenamefont {Berg}, \citenamefont {Kivelson},\ and\ \citenamefont {Raghu}}]{annurev}%
  \BibitemOpen
  \bibfield  {author} {\bibinfo {author} {\bibfnamefont {D.~P.}\ \bibnamefont {Arovas}}, \bibinfo {author} {\bibfnamefont {E.}~\bibnamefont {Berg}}, \bibinfo {author} {\bibfnamefont {S.~A.}\ \bibnamefont {Kivelson}},\ and\ \bibinfo {author} {\bibfnamefont {S.}~\bibnamefont {Raghu}},\ }\bibfield  {title} {\bibinfo {title} {{The Hubbard Model}},\ }\href {https://doi.org/https://doi.org/10.1146/annurev-conmatphys-031620-102024} {\bibfield  {journal} {\bibinfo  {journal} {"Annual Review of Condensed Matter Physics"}\ }\textbf {\bibinfo {volume} {13}},\ \bibinfo {pages} {239} (\bibinfo {year} {2022})}\BibitemShut {NoStop}%
\bibitem [{\citenamefont {Qin}\ \emph {et~al.}(2022)\citenamefont {Qin}, \citenamefont {Schäfer}, \citenamefont {Andergassen}, \citenamefont {Corboz},\ and\ \citenamefont {Gull}}]{gullrev}%
  \BibitemOpen
  \bibfield  {author} {\bibinfo {author} {\bibfnamefont {M.}~\bibnamefont {Qin}}, \bibinfo {author} {\bibfnamefont {T.}~\bibnamefont {Schäfer}}, \bibinfo {author} {\bibfnamefont {S.}~\bibnamefont {Andergassen}}, \bibinfo {author} {\bibfnamefont {P.}~\bibnamefont {Corboz}},\ and\ \bibinfo {author} {\bibfnamefont {E.}~\bibnamefont {Gull}},\ }\bibfield  {title} {\bibinfo {title} {{The Hubbard Model: A Computational Perspective}},\ }\href {https://doi.org/https://doi.org/10.1146/annurev-conmatphys-090921-033948} {\bibfield  {journal} {\bibinfo  {journal} {Annual Review of Condensed Matter Physics}\ }\textbf {\bibinfo {volume} {13}},\ \bibinfo {pages} {275} (\bibinfo {year} {2022})}\BibitemShut {NoStop}%
\bibitem [{\citenamefont {Fradkin}\ \emph {et~al.}(2015)\citenamefont {Fradkin}, \citenamefont {Kivelson},\ and\ \citenamefont {Tranquada}}]{Fradkin2015}%
  \BibitemOpen
  \bibfield  {author} {\bibinfo {author} {\bibfnamefont {E.}~\bibnamefont {Fradkin}}, \bibinfo {author} {\bibfnamefont {S.~A.}\ \bibnamefont {Kivelson}},\ and\ \bibinfo {author} {\bibfnamefont {J.~M.}\ \bibnamefont {Tranquada}},\ }\bibfield  {title} {\bibinfo {title} {{Colloquium: Theory of intertwined orders in high temperature superconductors}},\ }\href {https://doi.org/10.1103/RevModPhys.87.457} {\bibfield  {journal} {\bibinfo  {journal} {Rev. Mod. Phys.}\ }\textbf {\bibinfo {volume} {87}},\ \bibinfo {pages} {457} (\bibinfo {year} {2015})}\BibitemShut {NoStop}%
\bibitem [{\citenamefont {Chubukov}(1993)}]{chubukov}%
  \BibitemOpen
  \bibfield  {author} {\bibinfo {author} {\bibfnamefont {A.~V.}\ \bibnamefont {Chubukov}},\ }\bibfield  {title} {\bibinfo {title} {Kohn-luttinger effect and the instability of a two-dimensional repulsive fermi liquid at t=0},\ }\href {https://doi.org/10.1103/PhysRevB.48.1097} {\bibfield  {journal} {\bibinfo  {journal} {Phys. Rev. B}\ }\textbf {\bibinfo {volume} {48}},\ \bibinfo {pages} {1097} (\bibinfo {year} {1993})}\BibitemShut {NoStop}%
\bibitem [{\citenamefont {Raghu}\ \emph {et~al.}(2010)\citenamefont {Raghu}, \citenamefont {Kivelson},\ and\ \citenamefont {Scalapino}}]{Raghu2010}%
  \BibitemOpen
  \bibfield  {author} {\bibinfo {author} {\bibfnamefont {S.}~\bibnamefont {Raghu}}, \bibinfo {author} {\bibfnamefont {S.~A.}\ \bibnamefont {Kivelson}},\ and\ \bibinfo {author} {\bibfnamefont {D.~J.}\ \bibnamefont {Scalapino}},\ }\bibfield  {title} {\bibinfo {title} {{Superconductivity in the repulsive Hubbard model: An asymptotically exact weak-coupling solution}},\ }\href {https://doi.org/10.1103/PhysRevB.81.224505} {\bibfield  {journal} {\bibinfo  {journal} {Phys. Rev. B}\ }\textbf {\bibinfo {volume} {81}},\ \bibinfo {pages} {224505} (\bibinfo {year} {2010})}\BibitemShut {NoStop}%
\bibitem [{\citenamefont {Jiang}\ \emph {et~al.}(2018)\citenamefont {Jiang}, \citenamefont {Weng},\ and\ \citenamefont {Kivelson}}]{Jiang2018tJ}%
  \BibitemOpen
  \bibfield  {author} {\bibinfo {author} {\bibfnamefont {H.-C.}\ \bibnamefont {Jiang}}, \bibinfo {author} {\bibfnamefont {Z.-Y.}\ \bibnamefont {Weng}},\ and\ \bibinfo {author} {\bibfnamefont {S.~A.}\ \bibnamefont {Kivelson}},\ }\bibfield  {title} {\bibinfo {title} {{Superconductivity in the doped $\mathit{t}\ensuremath{-}\mathit{J}$ model: Results for four-leg cylinders}},\ }\href {https://doi.org/10.1103/PhysRevB.98.140505} {\bibfield  {journal} {\bibinfo  {journal} {Phys. Rev. B}\ }\textbf {\bibinfo {volume} {98}},\ \bibinfo {pages} {140505} (\bibinfo {year} {2018})}\BibitemShut {NoStop}%
\bibitem [{\citenamefont {Jiang}\ and\ \citenamefont {Devereaux}(2019)}]{Jiang2019Hub}%
  \BibitemOpen
  \bibfield  {author} {\bibinfo {author} {\bibfnamefont {H.-C.}\ \bibnamefont {Jiang}}\ and\ \bibinfo {author} {\bibfnamefont {T.~P.}\ \bibnamefont {Devereaux}},\ }\bibfield  {title} {\bibinfo {title} {{Superconductivity in the doped Hubbard model and its interplay with next-nearest hopping $t'$}},\ }\href {https://doi.org/10.1126/science.aal5304} {\bibfield  {journal} {\bibinfo  {journal} {Science}\ }\textbf {\bibinfo {volume} {365}},\ \bibinfo {pages} {1424} (\bibinfo {year} {2019})}\BibitemShut {NoStop}%
\bibitem [{\citenamefont {Jiang}\ \emph {et~al.}(2020)\citenamefont {Jiang}, \citenamefont {Zaanen}, \citenamefont {Devereaux},\ and\ \citenamefont {Jiang}}]{Jiang2020YF}%
  \BibitemOpen
  \bibfield  {author} {\bibinfo {author} {\bibfnamefont {Y.-F.}\ \bibnamefont {Jiang}}, \bibinfo {author} {\bibfnamefont {J.}~\bibnamefont {Zaanen}}, \bibinfo {author} {\bibfnamefont {T.~P.}\ \bibnamefont {Devereaux}},\ and\ \bibinfo {author} {\bibfnamefont {H.-C.}\ \bibnamefont {Jiang}},\ }\bibfield  {title} {\bibinfo {title} {{Ground state phase diagram of the doped Hubbard model on the four-leg cylinder}},\ }\href {https://doi.org/10.1103/PhysRevResearch.2.033073} {\bibfield  {journal} {\bibinfo  {journal} {Phys. Rev. Res.}\ }\textbf {\bibinfo {volume} {2}},\ \bibinfo {pages} {033073} (\bibinfo {year} {2020})}\BibitemShut {NoStop}%
\bibitem [{\citenamefont {Gong}\ \emph {et~al.}(2021)\citenamefont {Gong}, \citenamefont {Zhu},\ and\ \citenamefont {Sheng}}]{Gong2021}%
  \BibitemOpen
  \bibfield  {author} {\bibinfo {author} {\bibfnamefont {S.}~\bibnamefont {Gong}}, \bibinfo {author} {\bibfnamefont {W.}~\bibnamefont {Zhu}},\ and\ \bibinfo {author} {\bibfnamefont {D.~N.}\ \bibnamefont {Sheng}},\ }\bibfield  {title} {\bibinfo {title} {{Robust $d$-Wave Superconductivity in the Square-Lattice $t\text{\ensuremath{-}}J$ Model}},\ }\href {https://doi.org/10.1103/PhysRevLett.127.097003} {\bibfield  {journal} {\bibinfo  {journal} {Phys. Rev. Lett.}\ }\textbf {\bibinfo {volume} {127}},\ \bibinfo {pages} {097003} (\bibinfo {year} {2021})}\BibitemShut {NoStop}%
\bibitem [{\citenamefont {Lu}\ \emph {et~al.}(2024)\citenamefont {Lu}, \citenamefont {Chen}, \citenamefont {Zhu}, \citenamefont {Sheng},\ and\ \citenamefont {Gong}}]{Lu2024}%
  \BibitemOpen
  \bibfield  {author} {\bibinfo {author} {\bibfnamefont {X.}~\bibnamefont {Lu}}, \bibinfo {author} {\bibfnamefont {F.}~\bibnamefont {Chen}}, \bibinfo {author} {\bibfnamefont {W.}~\bibnamefont {Zhu}}, \bibinfo {author} {\bibfnamefont {D.~N.}\ \bibnamefont {Sheng}},\ and\ \bibinfo {author} {\bibfnamefont {S.-S.}\ \bibnamefont {Gong}},\ }\bibfield  {title} {\bibinfo {title} {{Emergent Superconductivity and Competing Charge Orders in Hole-Doped Square-Lattice $t-J$ Model}},\ }\href {https://doi.org/10.1103/PhysRevLett.132.066002} {\bibfield  {journal} {\bibinfo  {journal} {Phys. Rev. Lett.}\ }\textbf {\bibinfo {volume} {132}},\ \bibinfo {pages} {066002} (\bibinfo {year} {2024})}\BibitemShut {NoStop}%
\bibitem [{\citenamefont {Xu}\ \emph {et~al.}(2024{\natexlab{a}})\citenamefont {Xu}, \citenamefont {Chung}, \citenamefont {Qin}, \citenamefont {Schollwöck}, \citenamefont {White},\ and\ \citenamefont {Zhang}}]{Xu2024}%
  \BibitemOpen
  \bibfield  {author} {\bibinfo {author} {\bibfnamefont {H.}~\bibnamefont {Xu}}, \bibinfo {author} {\bibfnamefont {C.-M.}\ \bibnamefont {Chung}}, \bibinfo {author} {\bibfnamefont {M.}~\bibnamefont {Qin}}, \bibinfo {author} {\bibfnamefont {U.}~\bibnamefont {Schollwöck}}, \bibinfo {author} {\bibfnamefont {S.~R.}\ \bibnamefont {White}},\ and\ \bibinfo {author} {\bibfnamefont {S.}~\bibnamefont {Zhang}},\ }\bibfield  {title} {\bibinfo {title} {{Coexistence of superconductivity with partially filled stripes in the Hubbard model}},\ }\href {https://doi.org/10.1126/science.adh7691} {\bibfield  {journal} {\bibinfo  {journal} {Science}\ }\textbf {\bibinfo {volume} {384}},\ \bibinfo {pages} {eadh7691} (\bibinfo {year} {2024}{\natexlab{a}})}\BibitemShut {NoStop}%
\bibitem [{\citenamefont {Chen}\ \emph {et~al.}(2025)\citenamefont {Chen}, \citenamefont {Haldane},\ and\ \citenamefont {Sheng}}]{Chen2025}%
  \BibitemOpen
  \bibfield  {author} {\bibinfo {author} {\bibfnamefont {F.}~\bibnamefont {Chen}}, \bibinfo {author} {\bibfnamefont {F.~D.~M.}\ \bibnamefont {Haldane}},\ and\ \bibinfo {author} {\bibfnamefont {D.~N.}\ \bibnamefont {Sheng}},\ }\bibfield  {title} {\bibinfo {title} {{Global phase diagram of D-wave superconductivity in the square-lattice $t-J$ model}},\ }\href {https://doi.org/10.1073/pnas.2420963122} {\bibfield  {journal} {\bibinfo  {journal} {Proceedings of the National Academy of Sciences}\ }\textbf {\bibinfo {volume} {122}},\ \bibinfo {pages} {e2420963122} (\bibinfo {year} {2025})}\BibitemShut {NoStop}%
\bibitem [{\citenamefont {Hager}\ \emph {et~al.}(2005)\citenamefont {Hager}, \citenamefont {Wellein}, \citenamefont {Jeckelmann},\ and\ \citenamefont {Fehske}}]{Hager2005}%
  \BibitemOpen
  \bibfield  {author} {\bibinfo {author} {\bibfnamefont {G.}~\bibnamefont {Hager}}, \bibinfo {author} {\bibfnamefont {G.}~\bibnamefont {Wellein}}, \bibinfo {author} {\bibfnamefont {E.}~\bibnamefont {Jeckelmann}},\ and\ \bibinfo {author} {\bibfnamefont {H.}~\bibnamefont {Fehske}},\ }\bibfield  {title} {\bibinfo {title} {{Stripe formation in doped Hubbard ladders}},\ }\href {https://doi.org/10.1103/PhysRevB.71.075108} {\bibfield  {journal} {\bibinfo  {journal} {Phys. Rev. B}\ }\textbf {\bibinfo {volume} {71}},\ \bibinfo {pages} {075108} (\bibinfo {year} {2005})}\BibitemShut {NoStop}%
\bibitem [{\citenamefont {Ponsioen}\ \emph {et~al.}(2019)\citenamefont {Ponsioen}, \citenamefont {Chung},\ and\ \citenamefont {Corboz}}]{Ponsioen2019}%
  \BibitemOpen
  \bibfield  {author} {\bibinfo {author} {\bibfnamefont {B.}~\bibnamefont {Ponsioen}}, \bibinfo {author} {\bibfnamefont {S.~S.}\ \bibnamefont {Chung}},\ and\ \bibinfo {author} {\bibfnamefont {P.}~\bibnamefont {Corboz}},\ }\bibfield  {title} {\bibinfo {title} {{Period 4 stripe in the extended two-dimensional Hubbard model}},\ }\href {https://doi.org/10.1103/PhysRevB.100.195141} {\bibfield  {journal} {\bibinfo  {journal} {Phys. Rev. B}\ }\textbf {\bibinfo {volume} {100}},\ \bibinfo {pages} {195141} (\bibinfo {year} {2019})}\BibitemShut {NoStop}%
\bibitem [{\citenamefont {Qin}\ \emph {et~al.}(2020)\citenamefont {Qin}, \citenamefont {Chung}, \citenamefont {Shi}, \citenamefont {Vitali}, \citenamefont {Hubig}, \citenamefont {Schollw\"ock}, \citenamefont {White},\ and\ \citenamefont {Zhang}}]{Qin2020}%
  \BibitemOpen
  \bibfield  {author} {\bibinfo {author} {\bibfnamefont {M.}~\bibnamefont {Qin}}, \bibinfo {author} {\bibfnamefont {C.-M.}\ \bibnamefont {Chung}}, \bibinfo {author} {\bibfnamefont {H.}~\bibnamefont {Shi}}, \bibinfo {author} {\bibfnamefont {E.}~\bibnamefont {Vitali}}, \bibinfo {author} {\bibfnamefont {C.}~\bibnamefont {Hubig}}, \bibinfo {author} {\bibfnamefont {U.}~\bibnamefont {Schollw\"ock}}, \bibinfo {author} {\bibfnamefont {S.~R.}\ \bibnamefont {White}},\ and\ \bibinfo {author} {\bibfnamefont {S.}~\bibnamefont {Zhang}} (\bibinfo {collaboration} {Simons Collaboration on the Many-Electron Problem}),\ }\bibfield  {title} {\bibinfo {title} {{Absence of Superconductivity in the Pure Two-Dimensional Hubbard Model}},\ }\href {https://doi.org/10.1103/PhysRevX.10.031016} {\bibfield  {journal} {\bibinfo  {journal} {Phys. Rev. X}\ }\textbf {\bibinfo {volume} {10}},\ \bibinfo {pages} {031016} (\bibinfo {year} {2020})}\BibitemShut {NoStop}%
\bibitem [{\citenamefont {Jiang}\ \emph {et~al.}(2021)\citenamefont {Jiang}, \citenamefont {Scalapino},\ and\ \citenamefont {White}}]{Jiang2021}%
  \BibitemOpen
  \bibfield  {author} {\bibinfo {author} {\bibfnamefont {S.}~\bibnamefont {Jiang}}, \bibinfo {author} {\bibfnamefont {D.~J.}\ \bibnamefont {Scalapino}},\ and\ \bibinfo {author} {\bibfnamefont {S.~R.}\ \bibnamefont {White}},\ }\bibfield  {title} {\bibinfo {title} {{Ground-state phase diagram of the $t-t'-J$ model}},\ }\href {https://doi.org/10.1073/pnas.2109978118} {\bibfield  {journal} {\bibinfo  {journal} {Proceedings of the National Academy of Sciences}\ }\textbf {\bibinfo {volume} {118}},\ \bibinfo {pages} {e2109978118} (\bibinfo {year} {2021})}\BibitemShut {NoStop}%
\bibitem [{\citenamefont {Jiang}\ \emph {et~al.}(2022)\citenamefont {Jiang}, \citenamefont {Scalapino},\ and\ \citenamefont {White}}]{Jiang2022}%
  \BibitemOpen
  \bibfield  {author} {\bibinfo {author} {\bibfnamefont {S.}~\bibnamefont {Jiang}}, \bibinfo {author} {\bibfnamefont {D.~J.}\ \bibnamefont {Scalapino}},\ and\ \bibinfo {author} {\bibfnamefont {S.~R.}\ \bibnamefont {White}},\ }\bibfield  {title} {\bibinfo {title} {{Pairing properties of the $t\text{\ensuremath{-}}{t}^{\ensuremath{'}}\text{\ensuremath{-}}{t}^{\ensuremath{''}}\text{\ensuremath{-}}J$ model}},\ }\href {https://doi.org/10.1103/PhysRevB.106.174507} {\bibfield  {journal} {\bibinfo  {journal} {Phys. Rev. B}\ }\textbf {\bibinfo {volume} {106}},\ \bibinfo {pages} {174507} (\bibinfo {year} {2022})}\BibitemShut {NoStop}%
\bibitem [{\citenamefont {Jiang}\ \emph {et~al.}(2024)\citenamefont {Jiang}, \citenamefont {Devereaux},\ and\ \citenamefont {Jiang}}]{Jiang2024}%
  \BibitemOpen
  \bibfield  {author} {\bibinfo {author} {\bibfnamefont {Y.-F.}\ \bibnamefont {Jiang}}, \bibinfo {author} {\bibfnamefont {T.~P.}\ \bibnamefont {Devereaux}},\ and\ \bibinfo {author} {\bibfnamefont {H.-C.}\ \bibnamefont {Jiang}},\ }\bibfield  {title} {\bibinfo {title} {{Ground-state phase diagram and superconductivity of the doped Hubbard model on six-leg square cylinders}},\ }\href {https://doi.org/10.1103/PhysRevB.109.085121} {\bibfield  {journal} {\bibinfo  {journal} {Phys. Rev. B}\ }\textbf {\bibinfo {volume} {109}},\ \bibinfo {pages} {085121} (\bibinfo {year} {2024})}\BibitemShut {NoStop}%
\bibitem [{\citenamefont {Xu}\ \emph {et~al.}(2024{\natexlab{b}})\citenamefont {Xu}, \citenamefont {Jiang},\ and\ \citenamefont {Jiang}}]{Xu2024HubSix}%
  \BibitemOpen
  \bibfield  {author} {\bibinfo {author} {\bibfnamefont {Z.}~\bibnamefont {Xu}}, \bibinfo {author} {\bibfnamefont {H.-C.}\ \bibnamefont {Jiang}},\ and\ \bibinfo {author} {\bibfnamefont {Y.-F.}\ \bibnamefont {Jiang}},\ }\bibfield  {title} {\bibinfo {title} {{Superconductivity enhancement and particle-hole asymmetry: interplay with electron attraction in doped Hubbard model}},\ }\bibfield  {journal} {\bibinfo  {journal} {arXiv preprint}\ }\href {https://doi.org/10.48550/arXiv.2402.11255} {10.48550/arXiv.2402.11255} (\bibinfo {year} {2024}{\natexlab{b}}),\ \Eprint {https://arxiv.org/abs/2402.11255} {arXiv:2402.11255} \BibitemShut {NoStop}%
\bibitem [{\citenamefont {Liu}\ \emph {et~al.}(2025)\citenamefont {Liu}, \citenamefont {Zhai}, \citenamefont {Peng}, \citenamefont {Gu},\ and\ \citenamefont {Chan}}]{Liu2025}%
  \BibitemOpen
  \bibfield  {author} {\bibinfo {author} {\bibfnamefont {W.-Y.}\ \bibnamefont {Liu}}, \bibinfo {author} {\bibfnamefont {H.}~\bibnamefont {Zhai}}, \bibinfo {author} {\bibfnamefont {R.}~\bibnamefont {Peng}}, \bibinfo {author} {\bibfnamefont {Z.-C.}\ \bibnamefont {Gu}},\ and\ \bibinfo {author} {\bibfnamefont {G.~K.-L.}\ \bibnamefont {Chan}},\ }\bibfield  {title} {\bibinfo {title} {{Accurate Simulation of the Hubbard Model with Finite Fermionic Projected Entangled Pair States}},\ }\href {https://doi.org/10.1103/r4q9-4yvj} {\bibfield  {journal} {\bibinfo  {journal} {Phys. Rev. Lett.}\ }\textbf {\bibinfo {volume} {134}},\ \bibinfo {pages} {256502} (\bibinfo {year} {2025})}\BibitemShut {NoStop}%
\bibitem [{\citenamefont {Xu}\ \emph {et~al.}(2025)\citenamefont {Xu}, \citenamefont {Liu},\ and\ \citenamefont {Jiang}}]{Xu2025Dia}%
  \BibitemOpen
  \bibfield  {author} {\bibinfo {author} {\bibfnamefont {Z.}~\bibnamefont {Xu}}, \bibinfo {author} {\bibfnamefont {G.-X.}\ \bibnamefont {Liu}},\ and\ \bibinfo {author} {\bibfnamefont {Y.-F.}\ \bibnamefont {Jiang}},\ }\bibfield  {title} {\bibinfo {title} {{Stripes, pair density wave, and holon Wigner crystal in single-band Hubbard model on diagonal square lattice}},\ }\bibfield  {journal} {\bibinfo  {journal} {arXiv preprint}\ }\href {https://doi.org/10.48550/arXiv.2409.18833} {10.48550/arXiv.2409.18833} (\bibinfo {year} {2025}),\ \Eprint {https://arxiv.org/abs/2409.18833} {arXiv:2409.18833} \BibitemShut {NoStop}%
\bibitem [{\citenamefont {Wang}\ \emph {et~al.}(2025)\citenamefont {Wang}, \citenamefont {Jiang},\ and\ \citenamefont {Yao}}]{Wang2025EP}%
  \BibitemOpen
  \bibfield  {author} {\bibinfo {author} {\bibfnamefont {H.-X.}\ \bibnamefont {Wang}}, \bibinfo {author} {\bibfnamefont {Y.-F.}\ \bibnamefont {Jiang}},\ and\ \bibinfo {author} {\bibfnamefont {H.}~\bibnamefont {Yao}},\ }\bibfield  {title} {\bibinfo {title} {{Robust d-wave superconductivity from the Su-Schrieffer-Heeger-Hubbard model: possible route to high-temperature superconductivity}},\ }\href {https://doi.org/https://doi.org/10.1016/j.scib.2025.04.055} {\bibfield  {journal} {\bibinfo  {journal} {Science Bulletin}\ }\textbf {\bibinfo {volume} {70}},\ \bibinfo {pages} {2260} (\bibinfo {year} {2025})}\BibitemShut {NoStop}%
\bibitem [{\citenamefont {Fernandes}\ \emph {et~al.}(2019)\citenamefont {Fernandes}, \citenamefont {Orth},\ and\ \citenamefont {Schmalian}}]{Fernandes2019}%
  \BibitemOpen
  \bibfield  {author} {\bibinfo {author} {\bibfnamefont {R.~M.}\ \bibnamefont {Fernandes}}, \bibinfo {author} {\bibfnamefont {P.~P.}\ \bibnamefont {Orth}},\ and\ \bibinfo {author} {\bibfnamefont {J.}~\bibnamefont {Schmalian}},\ }\bibfield  {title} {\bibinfo {title} {{Intertwined Vestigial Order in Quantum Materials: Nematicity and Beyond}},\ }\href {https://doi.org/https://doi.org/10.1146/annurev-conmatphys-031218-013200} {\bibfield  {journal} {\bibinfo  {journal} {Annual Review of Condensed Matter Physics}\ }\textbf {\bibinfo {volume} {10}},\ \bibinfo {pages} {133} (\bibinfo {year} {2019})}\BibitemShut {NoStop}%
\bibitem [{\citenamefont {Mai}\ \emph {et~al.}(2022)\citenamefont {Mai}, \citenamefont {Karakuzu}, \citenamefont {Balduzzi}, \citenamefont {Johnston},\ and\ \citenamefont {Maier}}]{Mai2022}%
  \BibitemOpen
  \bibfield  {author} {\bibinfo {author} {\bibfnamefont {P.}~\bibnamefont {Mai}}, \bibinfo {author} {\bibfnamefont {S.}~\bibnamefont {Karakuzu}}, \bibinfo {author} {\bibfnamefont {G.}~\bibnamefont {Balduzzi}}, \bibinfo {author} {\bibfnamefont {S.}~\bibnamefont {Johnston}},\ and\ \bibinfo {author} {\bibfnamefont {T.~A.}\ \bibnamefont {Maier}},\ }\bibfield  {title} {\bibinfo {title} {{Intertwined spin, charge, and pair correlations in the two-dimensional Hubbard model in the thermodynamic limit}},\ }\href {https://doi.org/10.1073/pnas.2112806119} {\bibfield  {journal} {\bibinfo  {journal} {Proceedings of the National Academy of Sciences}\ }\textbf {\bibinfo {volume} {119}},\ \bibinfo {pages} {e2112806119} (\bibinfo {year} {2022})}\BibitemShut {NoStop}%
\bibitem [{\citenamefont {Corboz}\ \emph {et~al.}(2014)\citenamefont {Corboz}, \citenamefont {Rice},\ and\ \citenamefont {Troyer}}]{Corboz2014}%
  \BibitemOpen
  \bibfield  {author} {\bibinfo {author} {\bibfnamefont {P.}~\bibnamefont {Corboz}}, \bibinfo {author} {\bibfnamefont {T.~M.}\ \bibnamefont {Rice}},\ and\ \bibinfo {author} {\bibfnamefont {M.}~\bibnamefont {Troyer}},\ }\bibfield  {title} {\bibinfo {title} {{Competing States in the $t$-$J$ Model: Uniform $d$-Wave State versus Stripe State}},\ }\href {https://doi.org/10.1103/PhysRevLett.113.046402} {\bibfield  {journal} {\bibinfo  {journal} {Phys. Rev. Lett.}\ }\textbf {\bibinfo {volume} {113}},\ \bibinfo {pages} {046402} (\bibinfo {year} {2014})}\BibitemShut {NoStop}%
\bibitem [{\citenamefont {Zheng}\ \emph {et~al.}(2017)\citenamefont {Zheng}, \citenamefont {Chung}, \citenamefont {Corboz}, \citenamefont {Ehlers}, \citenamefont {Qin}, \citenamefont {Noack}, \citenamefont {Shi}, \citenamefont {White}, \citenamefont {Zhang},\ and\ \citenamefont {Chan}}]{Zheng2017}%
  \BibitemOpen
  \bibfield  {author} {\bibinfo {author} {\bibfnamefont {B.-X.}\ \bibnamefont {Zheng}}, \bibinfo {author} {\bibfnamefont {C.-M.}\ \bibnamefont {Chung}}, \bibinfo {author} {\bibfnamefont {P.}~\bibnamefont {Corboz}}, \bibinfo {author} {\bibfnamefont {G.}~\bibnamefont {Ehlers}}, \bibinfo {author} {\bibfnamefont {M.-P.}\ \bibnamefont {Qin}}, \bibinfo {author} {\bibfnamefont {R.~M.}\ \bibnamefont {Noack}}, \bibinfo {author} {\bibfnamefont {H.}~\bibnamefont {Shi}}, \bibinfo {author} {\bibfnamefont {S.~R.}\ \bibnamefont {White}}, \bibinfo {author} {\bibfnamefont {S.}~\bibnamefont {Zhang}},\ and\ \bibinfo {author} {\bibfnamefont {G.~K.-L.}\ \bibnamefont {Chan}},\ }\bibfield  {title} {\bibinfo {title} {{Stripe order in the underdoped region of the two-dimensional Hubbard model}},\ }\href {https://doi.org/10.1126/science.aam7127} {\bibfield  {journal} {\bibinfo  {journal} {Science}\ }\textbf {\bibinfo {volume} {358}},\ \bibinfo {pages} {1155} (\bibinfo {year} {2017})}\BibitemShut {NoStop}%
\bibitem [{\citenamefont {Zhang}\ \emph {et~al.}(2025)\citenamefont {Zhang}, \citenamefont {Li}, \citenamefont {Nikolaidou},\ and\ \citenamefont {von Delft}}]{Zhang2025}%
  \BibitemOpen
  \bibfield  {author} {\bibinfo {author} {\bibfnamefont {C.}~\bibnamefont {Zhang}}, \bibinfo {author} {\bibfnamefont {J.-W.}\ \bibnamefont {Li}}, \bibinfo {author} {\bibfnamefont {D.}~\bibnamefont {Nikolaidou}},\ and\ \bibinfo {author} {\bibfnamefont {J.}~\bibnamefont {von Delft}},\ }\bibfield  {title} {\bibinfo {title} {{Frustration-Induced Superconductivity in the $t\text{\ensuremath{-}}{t}^{\ensuremath{'}}$ Hubbard Model}},\ }\href {https://doi.org/10.1103/PhysRevLett.134.116502} {\bibfield  {journal} {\bibinfo  {journal} {Phys. Rev. Lett.}\ }\textbf {\bibinfo {volume} {134}},\ \bibinfo {pages} {116502} (\bibinfo {year} {2025})}\BibitemShut {NoStop}%
\bibitem [{\citenamefont {Wang}\ and\ \citenamefont {Devereaux}(2025)}]{Wang2025}%
  \BibitemOpen
  \bibfield  {author} {\bibinfo {author} {\bibfnamefont {W.~O.}\ \bibnamefont {Wang}}\ and\ \bibinfo {author} {\bibfnamefont {T.~P.}\ \bibnamefont {Devereaux}},\ }\href {https://arxiv.org/abs/2510.16616} {\bibinfo {title} {{Finite-temperature signatures of underlying superconductivity in the electron-doped Hubbard model}}} (\bibinfo {year} {2025}),\ \Eprint {https://arxiv.org/abs/2510.16616} {arXiv:2510.16616 [cond-mat.supr-con]} \BibitemShut {NoStop}%
\bibitem [{\citenamefont {White}(1992)}]{White1992}%
  \BibitemOpen
  \bibfield  {author} {\bibinfo {author} {\bibfnamefont {S.~R.}\ \bibnamefont {White}},\ }\bibfield  {title} {\bibinfo {title} {{Density matrix formulation for quantum renormalization groups}},\ }\href@noop {} {\bibfield  {journal} {\bibinfo  {journal} {Phys. Rev. Lett.}\ }\textbf {\bibinfo {volume} {69}},\ \bibinfo {pages} {2863} (\bibinfo {year} {1992})}\BibitemShut {NoStop}%
\bibitem [{\citenamefont {White}(1993)}]{White1993}%
  \BibitemOpen
  \bibfield  {author} {\bibinfo {author} {\bibfnamefont {S.~R.}\ \bibnamefont {White}},\ }\bibfield  {title} {\bibinfo {title} {{Density-matrix algorithms for quantum renormalization groups}},\ }\href {https://doi.org/10.1103/PhysRevB.48.10345} {\bibfield  {journal} {\bibinfo  {journal} {Phys. Rev. B}\ }\textbf {\bibinfo {volume} {48}},\ \bibinfo {pages} {10345} (\bibinfo {year} {1993})}\BibitemShut {NoStop}%
\bibitem [{\citenamefont {McCulloch}\ and\ \citenamefont {Gulácsi}(2002)}]{McCulloch2002}%
  \BibitemOpen
  \bibfield  {author} {\bibinfo {author} {\bibfnamefont {I.~P.}\ \bibnamefont {McCulloch}}\ and\ \bibinfo {author} {\bibfnamefont {M.}~\bibnamefont {Gulácsi}},\ }\bibfield  {title} {\bibinfo {title} {{The non-Abelian density matrix renormalization group algorithm}},\ }\href {https://doi.org/10.1209/epl/i2002-00393-0} {\bibfield  {journal} {\bibinfo  {journal} {Europhysics Letters}\ }\textbf {\bibinfo {volume} {57}},\ \bibinfo {pages} {852} (\bibinfo {year} {2002})}\BibitemShut {NoStop}%
\bibitem [{\citenamefont {Gu}\ \emph {et~al.}(2025)\citenamefont {Gu}, \citenamefont {Li}, \citenamefont {Lin}, \citenamefont {Zhan}, \citenamefont {Li}, \citenamefont {Huang}, \citenamefont {He}, \citenamefont {Wu}, \citenamefont {Xiang}, \citenamefont {Qin}, \citenamefont {Wang},\ and\ \citenamefont {Lv}}]{Gu2025ML}%
  \BibitemOpen
  \bibfield  {author} {\bibinfo {author} {\bibfnamefont {Y.}~\bibnamefont {Gu}}, \bibinfo {author} {\bibfnamefont {W.}~\bibnamefont {Li}}, \bibinfo {author} {\bibfnamefont {H.}~\bibnamefont {Lin}}, \bibinfo {author} {\bibfnamefont {B.}~\bibnamefont {Zhan}}, \bibinfo {author} {\bibfnamefont {R.}~\bibnamefont {Li}}, \bibinfo {author} {\bibfnamefont {Y.}~\bibnamefont {Huang}}, \bibinfo {author} {\bibfnamefont {D.}~\bibnamefont {He}}, \bibinfo {author} {\bibfnamefont {Y.}~\bibnamefont {Wu}}, \bibinfo {author} {\bibfnamefont {T.}~\bibnamefont {Xiang}}, \bibinfo {author} {\bibfnamefont {M.}~\bibnamefont {Qin}}, \bibinfo {author} {\bibfnamefont {L.}~\bibnamefont {Wang}},\ and\ \bibinfo {author} {\bibfnamefont {D.}~\bibnamefont {Lv}},\ }\href {https://arxiv.org/abs/2507.02644} {\bibinfo {title} {{Solving the Hubbard model with Neural Quantum States}}} (\bibinfo {year} {2025}),\ \Eprint {https://arxiv.org/abs/2507.02644} {arXiv:2507.02644 [cond-mat.str-el]} \BibitemShut {NoStop}%
\bibitem [{\citenamefont {Huang}\ \emph {et~al.}(2021)\citenamefont {Huang}, \citenamefont {Wang}, \citenamefont {Ding}, \citenamefont {Liu}, \citenamefont {Liu}, \citenamefont {Huang}, \citenamefont {Moritz},\ and\ \citenamefont {Devereaux}}]{Huang2021}%
  \BibitemOpen
  \bibfield  {author} {\bibinfo {author} {\bibfnamefont {E.~W.}\ \bibnamefont {Huang}}, \bibinfo {author} {\bibfnamefont {W.~O.}\ \bibnamefont {Wang}}, \bibinfo {author} {\bibfnamefont {J.~K.}\ \bibnamefont {Ding}}, \bibinfo {author} {\bibfnamefont {T.}~\bibnamefont {Liu}}, \bibinfo {author} {\bibfnamefont {F.}~\bibnamefont {Liu}}, \bibinfo {author} {\bibfnamefont {X.-X.}\ \bibnamefont {Huang}}, \bibinfo {author} {\bibfnamefont {B.}~\bibnamefont {Moritz}},\ and\ \bibinfo {author} {\bibfnamefont {T.~P.}\ \bibnamefont {Devereaux}},\ }\bibfield  {title} {\bibinfo {title} {{Intertwined States at Finite Temperatures in the Hubbard Model}},\ }\href {https://doi.org/10.7566/JPSJ.90.111010} {\bibfield  {journal} {\bibinfo  {journal} {Journal of the Physical Society of Japan}\ }\textbf {\bibinfo {volume} {90}},\ \bibinfo {pages} {111010} (\bibinfo {year} {2021})},\ \Eprint {https://arxiv.org/abs/https://doi.org/10.7566/JPSJ.90.111010} {https://doi.org/10.7566/JPSJ.90.111010} \BibitemShut {NoStop}%
\bibitem [{\citenamefont {Wietek}\ \emph {et~al.}(2021)\citenamefont {Wietek}, \citenamefont {He}, \citenamefont {White}, \citenamefont {Georges},\ and\ \citenamefont {Stoudenmire}}]{Wietek2021}%
  \BibitemOpen
  \bibfield  {author} {\bibinfo {author} {\bibfnamefont {A.}~\bibnamefont {Wietek}}, \bibinfo {author} {\bibfnamefont {Y.-Y.}\ \bibnamefont {He}}, \bibinfo {author} {\bibfnamefont {S.~R.}\ \bibnamefont {White}}, \bibinfo {author} {\bibfnamefont {A.}~\bibnamefont {Georges}},\ and\ \bibinfo {author} {\bibfnamefont {E.~M.}\ \bibnamefont {Stoudenmire}},\ }\bibfield  {title} {\bibinfo {title} {{Stripes, Antiferromagnetism, and the Pseudogap in the Doped Hubbard Model at Finite Temperature}},\ }\href {https://doi.org/10.1103/PhysRevX.11.031007} {\bibfield  {journal} {\bibinfo  {journal} {Phys. Rev. X}\ }\textbf {\bibinfo {volume} {11}},\ \bibinfo {pages} {031007} (\bibinfo {year} {2021})}\BibitemShut {NoStop}%
\bibitem [{\citenamefont {Chen}\ \emph {et~al.}(2021)\citenamefont {Chen}, \citenamefont {Wang}, \citenamefont {Rebec}, \citenamefont {Jia}, \citenamefont {Hashimoto}, \citenamefont {Lu}, \citenamefont {Moritz}, \citenamefont {Moore}, \citenamefont {Devereaux},\ and\ \citenamefont {Shen}}]{Chen2021}%
  \BibitemOpen
  \bibfield  {author} {\bibinfo {author} {\bibfnamefont {Z.}~\bibnamefont {Chen}}, \bibinfo {author} {\bibfnamefont {Y.}~\bibnamefont {Wang}}, \bibinfo {author} {\bibfnamefont {S.~N.}\ \bibnamefont {Rebec}}, \bibinfo {author} {\bibfnamefont {T.}~\bibnamefont {Jia}}, \bibinfo {author} {\bibfnamefont {M.}~\bibnamefont {Hashimoto}}, \bibinfo {author} {\bibfnamefont {D.}~\bibnamefont {Lu}}, \bibinfo {author} {\bibfnamefont {B.}~\bibnamefont {Moritz}}, \bibinfo {author} {\bibfnamefont {R.~G.}\ \bibnamefont {Moore}}, \bibinfo {author} {\bibfnamefont {T.~P.}\ \bibnamefont {Devereaux}},\ and\ \bibinfo {author} {\bibfnamefont {Z.-X.}\ \bibnamefont {Shen}},\ }\bibfield  {title} {\bibinfo {title} {{Anomalously strong near-neighbor attraction in doped 1D cuprate chains}},\ }\href {https://doi.org/10.1126/science.abf5174} {\bibfield  {journal} {\bibinfo  {journal} {Science}\ }\textbf {\bibinfo {volume} {373}},\ \bibinfo {pages} {1235} (\bibinfo {year} {2021})}\BibitemShut {NoStop}%
\bibitem [{\citenamefont {Peng}\ \emph {et~al.}(2023)\citenamefont {Peng}, \citenamefont {Wang}, \citenamefont {Wen}, \citenamefont {Lee}, \citenamefont {Devereaux},\ and\ \citenamefont {Jiang}}]{Peng2023}%
  \BibitemOpen
  \bibfield  {author} {\bibinfo {author} {\bibfnamefont {C.}~\bibnamefont {Peng}}, \bibinfo {author} {\bibfnamefont {Y.}~\bibnamefont {Wang}}, \bibinfo {author} {\bibfnamefont {J.}~\bibnamefont {Wen}}, \bibinfo {author} {\bibfnamefont {Y.~S.}\ \bibnamefont {Lee}}, \bibinfo {author} {\bibfnamefont {T.~P.}\ \bibnamefont {Devereaux}},\ and\ \bibinfo {author} {\bibfnamefont {H.-C.}\ \bibnamefont {Jiang}},\ }\bibfield  {title} {\bibinfo {title} {{Enhanced superconductivity by near-neighbor attraction in the doped extended Hubbard model}},\ }\href {https://doi.org/10.1103/PhysRevB.107.L201102} {\bibfield  {journal} {\bibinfo  {journal} {Phys. Rev. B}\ }\textbf {\bibinfo {volume} {107}},\ \bibinfo {pages} {L201102} (\bibinfo {year} {2023})}\BibitemShut {NoStop}%
\bibitem [{\citenamefont {Emery}(1987)}]{Emery1987}%
  \BibitemOpen
  \bibfield  {author} {\bibinfo {author} {\bibfnamefont {V.~J.}\ \bibnamefont {Emery}},\ }\bibfield  {title} {\bibinfo {title} {{Theory of high-${\mathrm{T}}_{\mathrm{c}}$ superconductivity in oxides}},\ }\href {https://doi.org/10.1103/PhysRevLett.58.2794} {\bibfield  {journal} {\bibinfo  {journal} {Phys. Rev. Lett.}\ }\textbf {\bibinfo {volume} {58}},\ \bibinfo {pages} {2794} (\bibinfo {year} {1987})}\BibitemShut {NoStop}%
\bibitem [{\citenamefont {Jiang}\ and\ \citenamefont {Kivelson}(2022)}]{Jiang2022Stripe}%
  \BibitemOpen
  \bibfield  {author} {\bibinfo {author} {\bibfnamefont {H.-C.}\ \bibnamefont {Jiang}}\ and\ \bibinfo {author} {\bibfnamefont {S.~A.}\ \bibnamefont {Kivelson}},\ }\bibfield  {title} {\bibinfo {title} {{Stripe order enhanced superconductivity in the Hubbard model}},\ }\href {https://doi.org/10.1073/pnas.2109406119} {\bibfield  {journal} {\bibinfo  {journal} {Proceedings of the National Academy of Sciences}\ }\textbf {\bibinfo {volume} {119}},\ \bibinfo {pages} {e2109406119} (\bibinfo {year} {2022})}\BibitemShut {NoStop}%
\bibitem [{Note1()}]{Note1}%
  \BibitemOpen
  \bibinfo {note} {\protect \href {https://doi.org/10.6084/m9.figshare.30689384}{Figshare dataset: 10.6084/m9.figshare.30689384}}\BibitemShut {NoStop}%
\end{thebibliography}
%

\clearpage
\newpage

\renewcommand{\thefigure}{S\arabic{figure}}
\setcounter{figure}{0}
\renewcommand{\theequation}{S\arabic{equation}}
\renewcommand{\thesection}{A\arabic{section}}
\setcounter{equation}{0}
\setcounter{page}{1}
\setcounter{table}{0}
\setcounter{section}{0}
\renewcommand{\thetable}{S\arabic{table}}

\begin{center}
\noindent {\large {\bf Supplemental Material}}
\end{center}
In the Supplemental Material, we present additional numerical results that further support the conclusions of the main text. Section~\ref{SM:Detail} provides detailed information on our DMRG calculations and demonstrates the reliability of the simulations. Section~\ref{SM:Energy} includes representative ground-state energies and describes the procedures used to determine the ground-state phase diagrams, including those shown in Fig.~\ref{Fig:Phase} of the main text. In Section~\ref{SM:PhaseBC}, we present the phase diagrams obtained under both PBC and ABC. Finally, Section~\ref{SM:Results} contains additional results for various correlation functions beyond those discussed in the main text.

\section{More numerical details}\label{SM:Detail} 
To ensure the reliability of our numerical results, we have carefully examined the robustness of our DMRG calculations for each set of parameters including system size, boundary conditions and model parameters. These examinations include using different states as starting points for the calculations, such as random initial states, states prepared under various pinning fields, and states obtained from simulations with different parameter settings. As this study focuses on finite-size systems, the true ground state is expected to preserve all continuous symmetries of the model Hamiltonian, including both the $U(1)$ charge-conservation symmetry and the $SU(2)$ spin-rotational symmetry. Preserving other symmetries - such as the translation symmetry defined with respect to the emergent charge density wave unil cell - is also important, as they have been found to significantly improve the convergence of the simulations.

\begin{figure}
  \includegraphics[width=1.0\linewidth]{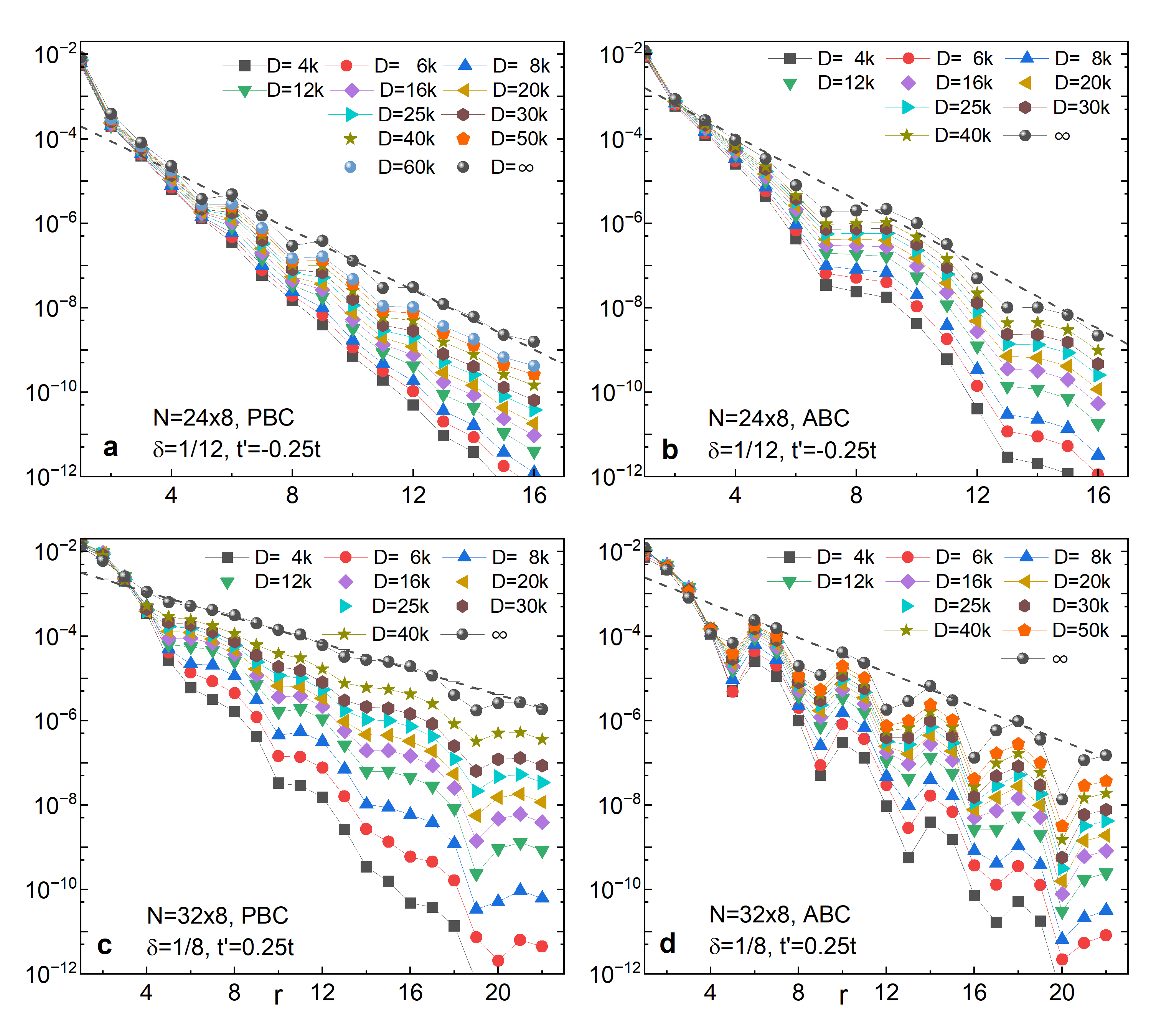}
\caption{(Color online) Convergence of superconducting correlations. Superconducting correlation $\Phi_{yy}(r)$ for (a-b) an $N=24 \times 8$ cylinder at $\delta=1/12$ with $t'=-0.25t$ under PBC and ABC, and for (c–d) an $N=32\times 8$ cylinder at $\delta=1/8$ with $t'=0.25t$ under the same boundary conditions. Results are presented for a range of $SU(2)$ multiplet bond dimensions $D$, together with their extrapolation to the limit $D=\infty$. All panels are plotted on semi-logarithmic scales, where $r$ is the separation between two Cooper pairs along the $\hat{x}$ direction. The black dashed lines indicate exponential fits of the form $\Phi_{yy}(r)\sim e^{-r/\xi_{SC}}$.}
\label{FigS:SCM}
\end{figure}

We have further assessed the convergence of our DMRG simulations by examining the stability of various correlation functions while varying the bond dimension of the $SU(2)$ multiplets from $D=1024$ to $D=60000$ under both PBC and ABC. As a representative example, Fig.~\ref{FigS:SCM} presents the superconducting (SC) pair-field correlations $\Phi_{yy}(r)$ for $N=24\times 8$ cylinders at $\delta=1/12$ with $t'=-0.25t$, and for $N=32\times 8$ at $\delta=1/8$ with $t'=0.25t$. The extrapolated values of $\Phi_{yy}(r)$ in the limit $D=\infty$ are obtained by fitting a second-order polynomial to four or five data points corresponding to the largest bond dimensions. Here, $r$ is the distance between two Cooper pairs along the $\hat{x}$ direction. To reduce boundary effects and local physics, several data points at small $r$ are omitted from the fitting procedure. As indicated by the black dashed lines, the SC correlations exhibit behavior consistent with an exponential decay $\Phi_{yy}(r) \sim e^{-r / \xi_{SC}}$ with the correlation length $\xi_{SC}$.

\begin{figure}[!th]
\centering
  \includegraphics[width=1\linewidth]{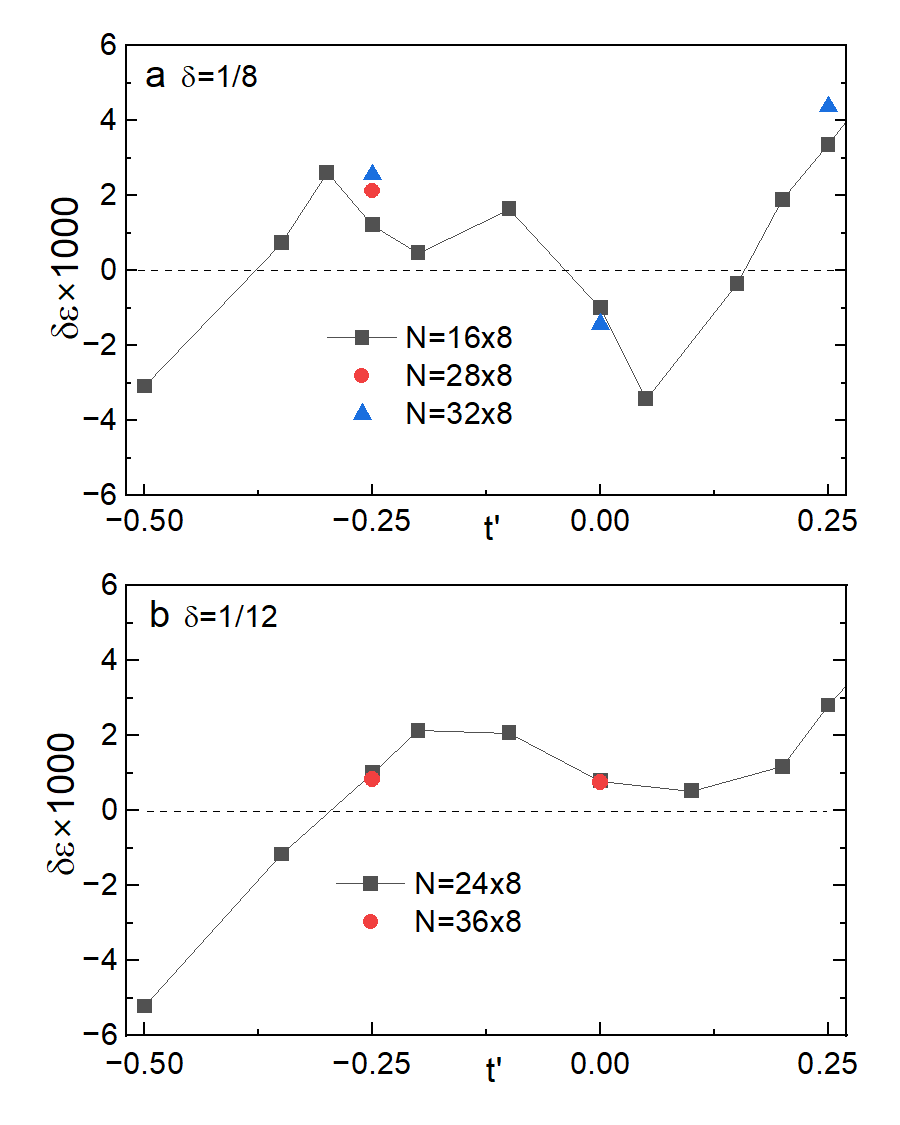}
\caption{Ground-state energy density difference between boundary conditions. The energy density difference $\delta \varepsilon$ for 8-leg Hubbard cylinders is defined such that $\delta \varepsilon > 0$ indicates the system with ABC has a lower energy, while $\delta \varepsilon < 0$ indicates the system with PBC has a lower energy. Panel (a) presents results at doping $\delta=1/8$ for cylinder lengths $L=16 - 32$, and panel (b) presents results at $\delta=1/12$ for $L=24 - 36$. Here, $U=12t$ and dashed lines are included as visual guides.
}\label{FigS:dE}
\end{figure}

\section{Ground state energy and energy difference}\label{SM:Energy} %
\begin{figure}[!th]
\centering
  \includegraphics[width=0.95\linewidth]{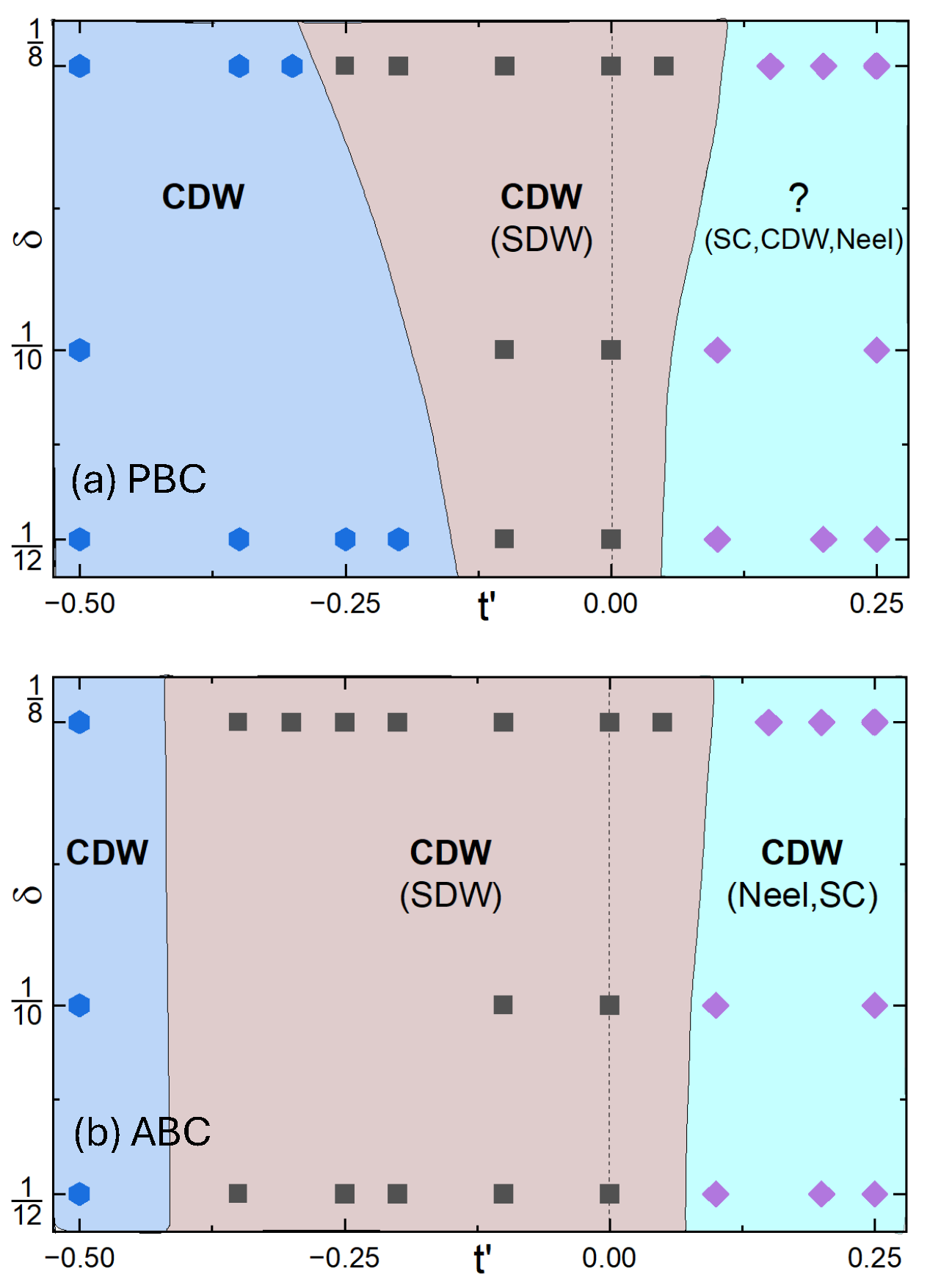}
\caption{{Ground state phase diagrams of the 8-leg Hubbard cylinders.} The ground state phase diagram of the Hubbard model on eight-leg square cylinders at $U=12t$ is shown as a function of $t'$ and doping $\delta$, for both (a) PBC and (b) ABC boundary conditions. Across most regions, the hierarchy of ordering tendencies is independent of boundary conditions. CDW order, typically of the half-filled stripe type and apparently long-range correlated except the region marked with a {\textbf{?}}), dominates most regimes labelled {\textbf{CDW}}; weaker or fluctuating CDW order is indicated when shown in parentheses. Regions with notable subdominant magnetic correlations — defined by a correlation length exceeding 4 lattice constants — are labeled (SDW) when locally incommensurate and (Néel) when commensurate. SC correlations are locally $d$-wave-like but generally decay exponentially; only in regions labeled (SC) does the SC correlation length exceed two lattice constants. Shaded regions serve as guides to the eye.
}\label{FigS:Phases}
\end{figure}

In this section, we describe the procedures used to determine the data points in the ground-state phase diagram shown in Fig.~\ref{Fig:Phase} of the main text, and the phase diagrams presented in Fig.~\ref{FigS:Phases} for PBC and ABC boundary conditions, respectively. For each system, we compute the ground-state with energy $E_0$ by ensuring that the convergence criteria required by different symmetries outlined in Sec.~\ref{SM:Detail} are satisfied. By retaining large bond dimensions in the DMRG calculations and by applying the standard extrapolation procedure using second-order polynomial function, we have extrapolated the ground-state energy to the limit $D=\infty$. The ground state energy density per site $E_0/N$ under different boundary conditions, i.e., $\varepsilon_{PBC}$ and $\varepsilon_{ABC}$, for different representative sets of parameters are provided in the Table. \ref{Tab:Sone}.

After extrapolating the ground-state energy to the limit $D=\infty$ for each set of parameters—specified by the system size $N$, doping level $\delta$, interaction strength $U$, and next-nearest-neighbor hopping $t'$ —we evaluate the energy per site difference between the two boundary conditions, defined as $\delta \varepsilon = \varepsilon_{PBC} - \varepsilon_{ABC}$. When $\delta \varepsilon > 0$, the system with ABC exhibits a lower ground-state energy, and the corresponding data points are represented by open symbols in Fig.~\ref{Fig:Phase}. Conversely, when $\delta \varepsilon < 0$, the system with PBC has a lower ground-state energy, and the corresponding data points are indicated by filled symbols. A more systematic analysis of $\delta \varepsilon$ at doping level $\delta=1/8$ is presented in Fig.~\ref{FigS:dE}, where results for different system sizes are compared. More numerical details are provided in Table~\ref{Tab:Sone}.

\begin{table*}[]
\centering
  \begin{tabular}{|c|c|c|c|c|c|c|c|c|c|c|c|c|}\hline
  \ $N$ \ & \ $\delta$ \ & \ $U$ \ & $t'$ & \ $\epsilon_{PBC}$ \ & \ $A_{PBC}$ \ & \ $\xi_{spin}$ \ & \ $\xi_{SC}$ \ & \ $\epsilon_{ABC}$ \ & \ $A_{ABC}$ \ & \ $\xi_{spin}$ \ & \ $\xi_{SC}$ \ & \ $\delta\varepsilon\times 10^{3}$ \ \\
\hline
  \ $32\times 8$ \ & 1/8 & \ 12$t$ \ & \ -0.25$t$ \ & \ -0.63572$t$ \ & 0.025 & 3.2 & 1.5 & \ -0.63829$t$ \ & 0.028 & 5.6 & 1.1 & +2.56$t$ \\
\hline
  $32\times 8$ & $1/8$ & 12$t$ & 0 & -0.63542$t$ & 0.033 & 4.2 & 1.4 & -0.63398$t$ & 0.023 & 5.9 & 1.4 & -1.43$t$\\
\hline
  $32\times 8$ & $1/8$ & 12$t$ & 0.25$t$ & -0.67483$t$ & 0.014 & 3.0 & 2.9 & -0.67921$t$ & 0.009 & 5.5 & 2.2 & +4.37$t$ \\
\hline
  $32\times 8$ \ & 1/8 & \ 8$t$ \ & \ -0.25$t$ \ & \ -0.74493$t$ \ & 0.025 & 3.5 & 1.5 & \ -0.74752$t$ \ & 0.024 & 6.7 & 1.2 & +2.60$t$ \\
\hline
  $32\times 8$ & $1/8$ & 8$t$ & 0 & -0.75551$t$ & 0.039 & 2.6 & 1.3 & -0.75184$t$ & 0.021 & 5.5 & 1.4 & -3.67$t$\\
\hline
  $24\times 8$ & \ $1/12$ \ & 12$t$ & -0.25$t$ & -0.55302$t$ & 0.019 & 1.5 & 1.3 & -0.55402$t$ & 0.025 & 6.9 & 1.2 & +1.00$t$\\
\hline
  $24\times 8$ & $1/12$ & 12$t$ & 0 & -0.54760$t$ & 0.024 & 5.9 & 1.4 & -0.54839$t$ & 0.030 & 7.4 & 1.2 & +0.79$t$\\
\hline
  $24\times 8$ & $1/12$ & 12$t$ & 0.25$t$ & -0.57388$t$ & 0.001 & 4.8 & 5.2 & -0.57669$t$ & 0.017 & 6.2 & 2.1 & +2.81$t$\\  
\hline
  $36\times 8$ & \ $1/12$ \ & 12$t$ & -0.25$t$ & -0.55487$t$ & 0.020 & 1.3 & 1.0 & -0.55631$t$ & 0.023 & 6.6 & 1.1 & +0.83$t$\\ 
\hline
  $36\times 8$ & $1/12$ & 12$t$ & 0 & -0.54994$t$ & 0.025 & 5.7 & 1.3 & -0.55069$t$ & 0.026 & 8.2 & 1.2 & +0.75$t$\\ \hline
\end{tabular}
\caption{Representative results for different sets of parameters, including system size $N$, doping concentration $\delta$, Coulomb repulsion $U$, second neighbor electron hopping $t'$ and boundary conditions (PBC and ABC). $\epsilon_{PBC}$ and $\epsilon_{ABC}$ are the ground state energy density per site of the system under PBC and ABC, respectively. $\delta \varepsilon = \epsilon_{PBC}-\epsilon_{ABC}$ is the difference in the ground-state energy density per site of the system with PBC- ABC - i.e. ABC has lower energy when $\delta \varepsilon >0$. $A_{PBC}$ and $A_{ABC}$ are CDW orders which are calculated using Eq.(7) in the middle half of the system. $\xi_{spin}$ and $\xi_{SC}$ are, respectively, the spin and SC correlation lengths.}
\label{Tab:Sone}
\end{table*}

\section{Ground state Phase diagram}\label{SM:PhaseBC}%
In addition to the ground-state phase diagram of the eight-leg Hubbard ladder presented in the main text, this section provides complementary ground state phase diagrams under PBC and ABC boundary conditions, shown in Fig.\ref{FigS:Phases}. Comparison with Fig. 1 in the main-text makes clear that all three phase diagrams are topologically similar, although quantitative differences are observed. In the regime $t'<0$ labeled \textbf{CDW}, the ground state is dominated by strong CDW order, with spin and SC correlations weak and exponentially decaying over one to two lattice spacings. In the middle region including $t'\approx 0$, CDW order remains dominant, SC correlations decay similarly, and short-range but relatively strong SDW correlations emerge, with correlation lengths typically exceeding four lattice spacings. 
When $t'>0$, while the ground state properties depend strongly on boundary conditions, the SC correlations are notably enhanced relative to the other regimes.

Differences between boundary conditions are most apparent for $t'>0$. Under ABC, the lower-energy ground states remain CDW-dominated, with strong antiferromagnetic spin-spin correlations (with spin-spin correlation lengths $\xi_s\geq 2$ lattice spacings) and moderately enhanced but still short-ranged SC correlations (with correlation lengths $\xi_{sc} \sim 2$ or larger). Under PBC, the ground-state energy is higher as shown in Fig.\ref{FigS:dE}, CDW order is suppressed, but SC correlations are significantly stronger, potentially becoming the dominant order at low doping. For example, at $\delta=1/12$ with $t'=0.25$, the SC correlation length can reach $\xi_{sc} \sim 5$.

\section{Correlation functions}\label{SM:Results} %
In this section, we present additional results for various correlation functions extrapolated to the limit $D=\infty$. These correlations include the single-particle Green’s function $G(r)=\langle \hat{c}^\dagger_{{\bf R}\sigma} \hat{c}_{{\bf R}+r\hat{x}}\rangle$, the spin–spin correlation function $F(r)$, and the spin-singlet SC correlation functions $\Phi_{xx}(r)$ and $\Phi_{yy}(r)$, evaluated for both $t'=\pm 0.25t$. Spin-triplet SC correlations were also computed but not shown here, but these were found to be significantly weaker than their spin-singlet counterparts, indicating that spin-triplet superconductivity is unlikely in this model.

\begin{figure}
  \includegraphics[width=1.0\linewidth]{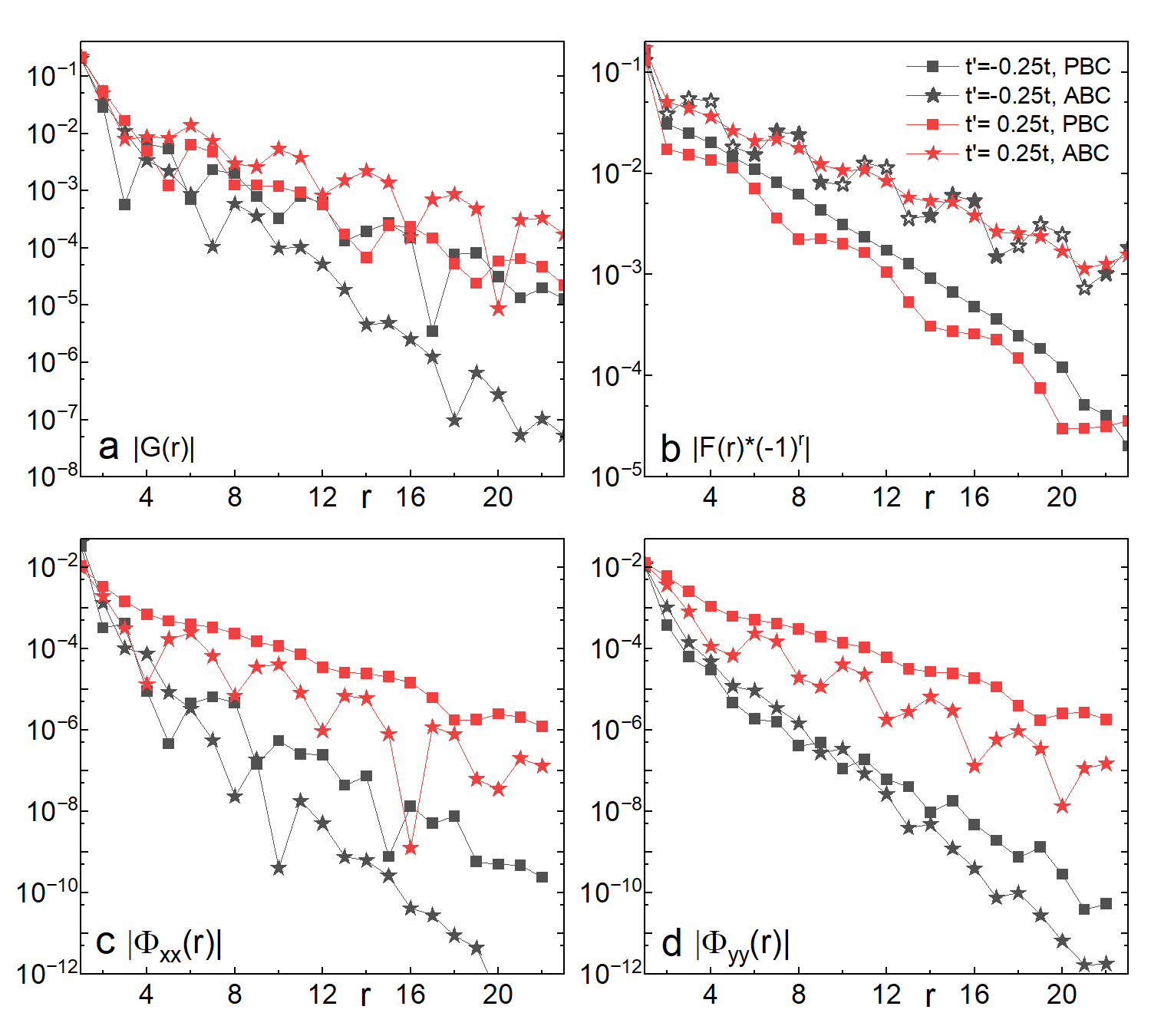}
\caption{(Color online) Correlation functions at 1/8 doping level. Correlation functions for the $N=32\times 8$ cylinder at $\delta=1/8$ and $t'=\pm 0.25t$ on semi-logarithmic scales as a function of distance $r$ along the cylinder: (a) single-particle Green’s function ($|G(r)|$), (b) spin–spin correlations $F(r)*(-1)^r$, and SC correlations (c) $\Phi_{xx}(r)$ and (d) $\Phi_{yy}(r)$. The spin correlations are locally N'eel-like (oscillating with period 1 along the $\hat{x}$ direction, so that $F(r)* (-1)^r$ has no sign change) for $t' > 0$, and locally SDW-like (oscillating with a $\delta$-dependent period) for $t'<0$. The sign structure of spin-spin correlations is indicated by the symbols: open symbols correspond to negative values, and filled symbols to positive values.}
\label{FigS:R8Cor}
\end{figure}

\begin{figure}
  \includegraphics[width=1.0\linewidth]{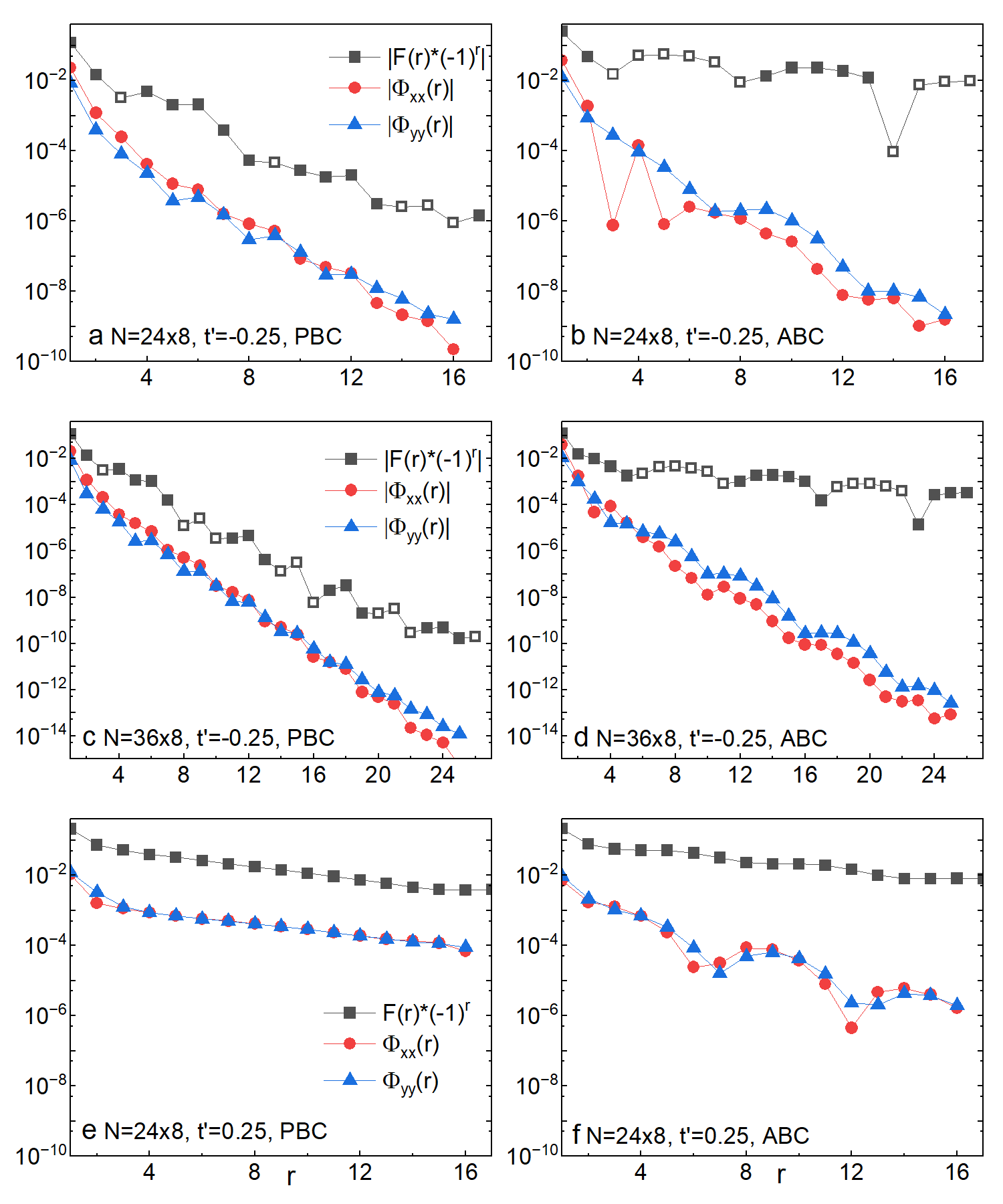}
\caption{(Color online) Correlation functions at 1/12 doping level. Correlation functions for the $N=24\times 8$ and $N=36\times 8$ cylinders at $\delta=1/12$ and $t'=\pm 0.25t$ are shown on semi-logarithmic scales as a function of distance $r$ along the cylinders, under both PBC and ABC. These include the spin–spin correlations $F(r)*(-1)^r$ and SC correlations $\Phi_{xx}(r)$ and $\Phi_{yy}(r)$. The spin correlations are locally N'eel-like (oscillating with period 1 along the $\hat{x}$ direction, so that $F(r)*(-1)^r$ exhibits no sign change) for $t'>0$, and locally SDW-like (oscillating with a $\delta$-dependent period) for $t'<0$. The sign structure of $F(r)*(-1)^r$ is indicated by the symbols: open symbols correspond to negative values, and filled symbols to positive values.}
\label{FigS:R12Cor}
\end{figure}

Figs~\ref{FigS:R8Cor} and \ref{FigS:R12Cor} show the correlations for $N=32\times 8$ at $\delta=1/8$ and for $N=24\times 8$ and $N=36\times 8$ at $\delta=1/12$, respectively, all plotted on semi-logarithmic scales. In every case, the correlation functions exhibit exponential decay, $F(r) \sim e^{-r/\xi_{spin}}$ and $\Phi(r) \sim e^{-r/\xi_{SC}}$, with the correlation lengths $\xi_{spin}$ and $\xi_{SC}$ listed in Table.~\ref{Tab:One} of the main text and Table.~\ref{Tab:Sone} in this Supplemental Material.

\end{document}